\documentclass[journal=enfuem,manuscript=article,hyperref]{achemso}
\usepackage{hyperref}
\usepackage{achemso}

\usepackage[version=4]{mhchem} % Formula subscripts using \ce{}
\usepackage[T1]{fontenc}
\usepackage{booktabs}
\usepackage[english]{babel}
\usepackage{csquotes}
\usepackage{textcomp}
\usepackage{amssymb}

\usepackage{graphicx}
\usepackage{caption,subcaption}
\usepackage[capitalise]{cleveref}

\usepackage{enumitem}
\usepackage{multicol}
\usepackage{siunitx}
\DeclareSIUnit\atm{atm}

\graphicspath{{./Figures/}}

\usepackage{lineno}
\usepackage[textsize=footnotesize,textwidth=2.25cm]{todonotes}

\author{Shane R.~Daly}
\affiliation[osu]
{School of Mechanical, Industrial, and Manufacturing Engineering\\
	Oregon State University, Corvallis, OR 97331, USA}

\author{Kyle E.~Niemeyer}
\affiliation[osu]{School of Mechanical, Industrial, and Manufacturing Engineering\\
	Oregon State University, Corvallis, OR 97331, USA}

\author{William J.~Cannella}
\affiliation[chev]
{Chevron Energy Technology Company Richmond, CA 94802, USA}

\author{Christopher L.~Hagen}
\email{chris.hagen@oregonstate.edu}
\phone{541-322-2061}
\fax{}
\affiliation[casc]
{School of Mechanical, Industrial, and Manufacturing Engineering\\
	Oregon State University--Cascades, Bend, OR 97703, USA}

\title[FACE gasoline surrogates formulated by an enhanced multivariate optimization framework]
{FACE gasoline surrogates formulated by an enhanced multivariate optimization framework}

\abbreviations{FACE, RON, MON, S, PCR, PIONA}

\begin{document}


\begin{abstract}

Design and optimization of higher efficiency, lower-emission internal combustion engines are highly dependent on fuel chemistry.
Resolving chemistry for complex fuels, like gasoline, is challenging. A solution is to study a fuel surrogate: a blend of a small number of well-characterized hydrocarbons to represent real fuels by emulating their thermophysical and chemical kinetics properties.
In the current study, an existing gasoline surrogate formulation algorithm is further enhanced by incorporating novel chemometric models.
These models use infrared spectra of hydrocarbon fuels to predict octane numbers, and are valid for a wide array of neat hydrocarbons and mixtures of such.
This work leverages 14 hydrocarbon species to form tailored surrogate palettes for the Fuels for Advanced Combustion Engine (FACE) gasolines, including candidate component species not previously considered: \textit{n}-pentane, 2-methylpentane, 1-pentene, cyclohexane, and \textit{o}-xylene. 
We evaluate the performance of ``full'' and ``reduced'' surrogates for the 10 fuels for advanced combustion engine (FACE) gasolines, containing between 8--12 and 4--7 components, respectively.
These surrogates match the target properties of the real fuels, on average, within \SI{5}{\percent}.
This close agreement demonstrates that the algorithm can design surrogates matching the wide array of target properties: octane numbers, density, hydrogen-to-carbon ratio, distillation characteristics, and proportions of carbon--carbon bond types. 
We also compare our surrogates to those available in literature (FACE gasolines A, C, F, G, I and J).
Our surrogates for these fuels, on average, better-match RON, MON, and distillation characteristics within 0.5\%, 0.7\%, and 0.9\%, respectively, with literature surrogates at 1.2\%, 1.1\%, and 1.8\% error.
However, our surrogates perform slightly worse for density, hydrogen-to-carbon ratio, and carbon--carbon bond types at errors of 3.3\%, 6.8\%, and 2.2\% with literature surrogates at 1.3\%, 2.3\%, and 1.9\%.
Overall, the approach demonstrated here offers a promising method to better design surrogates for gasoline-like fuels with a wide array of properties.
\end{abstract}
\clearpage

\section{Introduction}
\label{S:Intro}
 
Combustion of hydrocarbon-based fuels accounted for approximately \SI{84}{\percent} of US energy consumption in April 2017~\cite{EIAmonthly:jul17}, and some projections show this remaining at \SIrange{82}{84}{\percent} through 2050~\cite{EIAoutlook}.
Society continues to rely on internal combustion engines for transportation, commerce, and power generation.
Vehicle fuel economy has steadily improved over the years due to increasing compression ratios, downsizing, and turbocharging, resulting in increased thermal efficiency~\cite{heywood2009trends}; in addition, increased electric hybridization has helped reduce overall fuel consumption.
However, simultaneously evolving fuel compositions and combustion strategies challenge further improvements to ICE performance---both are being adapted to meet high efficiency, low emission government mandates. 
These recent governmental regulations, namely the Corporate Average Fuel Economy 2025 standards, propose reaching 54.5 mpg within the next eight years, although they are currently under review.  
Continued electric hybridization will help solve this challenge, but projections suggest personal and fleet hybrid vehicles will comprise only \SI{4.6}{\percent} of on-road vehicles in the United States by 2025~\cite{EIAoutlook2}. 
Reducing the environmental impact of these systems motivates research into further reduction of emissions and improvements in efficiency.

The chemical composition of a fuel significantly affects engine-out emissions and performance~\cite{dec2004isolating,Niemeyer2015a,Truedsson:2014ut,costa2010hydrous}.
Gasoline presents challenges to studying the influence of fuel composition since it contains hundreds of various hydrocarbon species~\cite{pitz2007development}. 
The methodology proposed in this work will accelerate the process of studying complex gasoline samples by formulating fuel surrogates.
Surrogates blend a small number of well-characterized hydrocarbons to represent real fuels (like refinery-grade gasolines) by emulating their thermophysical and chemical kinetics properties.
By representing these real fuels---comprised of near-continuous spectra of component hydrocarbons---as discrete mixtures of components, surrogates can be modeled with validated chemical kinetic models and enable simulations of combustion technology.
%It is not yet possible to directly simulate a gasoline fuel comprised of hundreds of hydrocarbons.  
Additionally, experiments with surrogate fuels can provide insight into fuel-composition effects on engine-out performance and emissions.
%(for example, Luo et al.~\cite{luo2012experimental}).
Experimental and computational research using surrogates designed with the proposed algorithm will help inform researchers to effects of fuel composition and thermochemical properties on internal combustion engine performance.

Our work builds upon the diesel surrogate formulation framework developed by Mueller et al.~\cite{Mueller2012}.
They demonstrated that physical properties and fuel performance metrics relevant to internal combustion engine performance---composition based on carbon nuclear magnetic resonance spectra, distillation curve, cetane number, and density---can be targeted to formulate surrogates that mimic the behavior of the real diesel fuel properties.
Their work helped overcome the challenge of automating the design of a mixture containing a small number of hydrocarbon species (eight, in this case) to broadly represent the performance of complex fuels.
Their method weights individual fuel properties and combines these into a single objective function, then designs a surrogate by minimizing this objective via changing the relative component amounts.

Ahmed et al.~\cite{Ahmed2015} extended the diesel surrogate formulation framework to gasoline fuels: Fuels for Advanced Combustion Engines (FACE) gasolines A and C. The FACE gasolines are a matrix of research gasolines designed by the Coordinating Research Council (CRC) and manufactured by ChevronPhillips Chemical Co.~\cite{Cannella:2014aa}.  Ahmed et al. efforts incorporated fuel performance metrics more relevant to gasoline fuels; e.g., research octane number (RON) in lieu of cetane number.
Later, Sarathy et al.~\cite{Sarathy2016} presented multiple surrogates for FACE gasolines F and G, using the method of Ahmed et al.~\cite{Ahmed2015} with additional target properties like motor octane number (MON).
Multiple surrogates were presented for FACE gasolines F and G based on species palette selection restrictions and a RON and MON calculation methodology. 
After which, Javed et al.~\cite{Javed2017} recently used the refined methodology of Sarathy et al.~\cite{Sarathy2016} to formulate surrogates for FACE I and J.
Shankar et al.~\cite{shankar2016primary} also formulated surrogates for low octane sensitivity gasolines FACE A, C, I and J, mainly to investigate the application of using primary reference fuels as surrogates specifically for pre-mixed, low-temperature combustion engine applications. 

\subsection{Octane model review}
\label{S:octane-review}

Here, we outline the various methodologies to develop surrogates whose RON and MON attempt to match those of the target gasoline in the aforementioned efforts~\cite{Ahmed2015,Sarathy2016, Javed2017}, while highlighting the challenges and areas for improvement.  Following that, we present an alternative approach to calculate RON and MON. 

Ahmed et al.~\cite{Ahmed2015} correlated RON to simulated constant-volume ignition delay times.
Their approach requires an autoignition simulation for every iteration in the surrogate formulation framework.
This step can be time-consuming, especially with detailed chemical models containing thousands of species, since the computational time increases exponentially with the number of component species considered for the surrogate---and hundreds of iterations may be required to formulate a surrogate.
Ahmed et al.\ also considered a simpler linear-by-mole correlation and formulated alternate surrogates based on this less-expensive computational approach~\cite{Ahmed2015}. 

Sarathy et al.~\cite{Sarathy2016} added octane sensitivity (S = RON-MON) as a target parameter when formulating surrogates for FACE gasolines F and G~\cite{Sarathy2016}.
They correlated octane sensitivity to the slope of the negative temperature coefficient (NTC) region of simulated constant volume ignition delays---a significant computational hurdle, compared with a single ignition delay calculation (see Mehl et al.~\cite{Mehl2011NTC} for more detail on this correlation).
To alleviate computational effort, Sarathy et al.\ also formed alternative surrogates for FACE gasolines F and G based on another linear-by-mole blending formula~\cite{Sarathy2016}.
The linear-by-mole equation, while simple, was developed and verified only for toluene reference fuels (TRFs: mixtures of toluene, \textit{n}-heptane, and isooctane).
Sarathy et al.\ extended the application to non-TRF mixtures by replacing toluene, \textit{n}-heptane, and isooctane as the aromatic, \textit{n}-paraffin, and isoparaffin species in the linear-by-mole equation with other species being considered.
This is to say, predicting RON or MON of a fuel mixture containing 1,2,4-trimethylbenzene, \textit{n}-butane, and 2-methylbutane, would be achieved by using the respective mole fractions directly in the equation originally developed for only TRF's.
As such, errors using the equation in this manner may be significant.
Despite this, Sarathy et al. did this to investigate an approach requiring minimal computational effort in comparison to the other, computationally-heavy method. 

Javed et al. replaced the TRF linear-by-mole relationship with the more detailed octane blending model of Ghosh et al.~\cite{Ghosh2006} in formulating FACE J. This relation accounts for non-linear blending effects at the level of the total paraffins, total olefins, total naphthenes, etc., but not at the individual molecule level~\cite{Ghosh2006}.  The Ghosh et al. model interaction parameters that represent blending effects between hydrocarbon classes were trained using gasoline samples.  After, the trained model was validated against other refinery-grade gasoline samples, and proved to produce excellent results.  While this octane model does return individual pure component octane ratings for neat hydrocarbons, it is not validated toward simpler surrogate mixtures where molecule-molecule interactions can be more prominent.

The RON and MON calculation approaches used in previous gasoline surrogate formulation efforts~\cite{Sarathy2014,Ahmed2015,Sarathy2016} are either not very accurate (the TRF linear-by-mole relationship), do not account for molecule-molecule octane blending, or computationally expensive (the NTC-sensitivity correlation based on multiple autoignition simulations).
In the current study, we explored improved options that require less computational effort and are capable of predicting quantities for a wide array of hydrocarbons.
Initially, we attempted to extend the correlation of simulated ignition delay approach of Ahmed et al.~\cite{Ahmed2015} to calculate MON, since they only developed a correlation for RON.  
The coefficient of fit for our MON model was sufficient, but the RON and MON models in combination could not accurately capture sensitivity.
More recently, Singh et al.~\cite{singh2017chemical} used regression tools to find the initial temperature and pressure conditions at which RON and MON best correlate with the simulated ignition delay times; it is applicable to mixtures of \emph{n}-heptane, isooctane, toluene, 1-hexene, and 1,2,4-trimethylbenzene.

We instead implemented novel models that correlate attenuated total reflectance, Fourier-transform infrared (ATR-FTIR) spectra of a fuel to RON and MON. 
These models alleviate the extensive computational effort of auto-ignition simulations---decoupling the need for complex chemical kinetic models---and provide accurate predictions for robust mixtures of various hydrocarbons, as well as more complex fuels like the FACE gasolines.
Daly et al.~\cite{Daly:2016} provide such a model developed through principal component regression (PCR), informed by neat hydrocarbons and low-component fuel mixtures.
The model can predict RON for various pure components and their mixtures, as well as the higher-complexity FACE gasolines.
The model takes a fuel's IR spectra, whether it is a pure component, mixture of pure components, or a refinery-grade gasoline sample, and predicts the RON. 
Here, we use this technique to create a separate MON correlation, and also improve the predictive capability of the original RON correlation provided by Daly et al.~\cite{Daly:2016}.  

The surrogate formulation algorithm, updated with the IR-octane models, is used to generate surrogates for the 10 FACE gasolines A--J. 
Table~\ref{T:FACE_gasolines} list the measured properties of these fuels~\cite{Cannella:2014aa}.
We next describe the IR-octane models and the surrogate formulation algorithm. 

\begin{table}[htbp]
\centering
\begin{tabular}{@{}l c c c c c c c c c c@{}}
\toprule
&\multicolumn{10}{c}{\small{FACE gasolines}} \\
\cmidrule{2-11}
Target property &A& B & C & D & E & F & G & H & I & J \\
\midrule
RON 	        &83.9 & 95.8 &  84.3  & 94.2 & 87.4 & 94.4 & 96.8 &	86.9 &	70.2  & 73.8 \\
MON		        &83.5 & 92.4 &  83.0  & 87.0 & 81.1 & 88.8 & 85.8 &	79.8 &	69.5  & 70.1 \\
Density [\si{\kilo\gram\per\meter^3}]	&685  & 697  &	690   & 743  & 725  & 707  & 760  &	759 &  697   & 742  \\
H/C ratio		&2.29 & 2.21 &	2.27  & 1.88 & 2.04 & 2.13 & 1.83 &	1.72 &	2.26  & 1.92 \\
\midrule
\% Volume distilled	& \multicolumn{10}{c}{Temperature [\si{\kelvin}]}  \\
\midrule
10			& 329 & 337 &   331	& 338 &335 & 346& 350& 334&343 &346 \\
20		    & 344 & 352 &   341	& 354 &342 & 351& 363& 349&354 &368 \\
30	        & 357 & 364 &   350 & 366 &348 & 357& 378& 363&359 &376 \\
40	        & 365 & 371 &	359	& 374 &354 & 363& 394& 373&362 &380 \\
50	        & 368 & 374 &	366	& 379 &359 & 370& 411& 382&364 &384 \\
60	        & 370 & 376 &	372	& 384 &365 & 376& 426& 390&366 &390 \\
70	        & 372 & 378 &	376	& 390 &371 & 382& 439& 401&368 &401 \\
80	        & 374 & 381 &	382	& 413 &379 & 387& 447& 417&371 &417 \\
\midrule
Carbon type	& \multicolumn{10}{c}{ Fractional \%}  \\
\midrule
1&	54.5&	56.4&	53.1&	41.3&	33.2&	43.2&	35.4&	27.5&	46.1& 32.8\\
2&	22.7&	15.2&	24.6&	18.4&	19.1&	18.2&	13&	22.4&	32.3&	32.8\\
3&	16.6&	16.6&	14.6&	5.2	&8.2&	12.1&	6.5&	3&	13.4&	4.5\\
4&	0.8&	0&	0	&0.2&	21.7&	11.5&	10.5&	9.5	&1.8&	1.1\\
5&	0.4	&0&	0	&0&	1&	0.1	&0.8&	0.8&	0.7&	0.6\\
6&	0&	0	&0&	0&	0&	0&	0&	0&	0&	0\\
7&	0.5&	4.1&	2&	20.7&	8.3&	5.3&	18.4&	24.8&	1.1	&18.1\\
8&	0.1	&1.8&	1.3	&9.3&	2.7	&2.3&	10.8&	9.1&	0.2	&9.9\\
9&	0&	0&	0&	0&	0&	0&	0&	0&	0&	0\\
10&	0&	0&	0	&0.2&	0.3&	0&	0.5&	0.5&	0&	0.1\\
11&	4.4&	5.8	&4.4&	4.7&	1.9	&4.2&	1.8&	0.5&	2.3&	0\\
12&	0&	0&	0&	0&	1.3&	1.5&	0.3	&0.3	&1&	0\\
13&	0&	0&	0&	0&	1.9&	1.5&	1.3&	1.1	&1&	0\\
14&	0&	0&	0&	0&	0.4&	0&	0.6	&0.6&	0&	0.1\\

\bottomrule
\end{tabular}
\caption{
FACE gasoline target properties~\cite{Cannella:2014aa}, including carbon bond (C--C) type relative amounts (rounded to nearest tenth); nomenclature for C--C is defined in the Methodology section. 
}
\label{T:FACE_gasolines}
\end{table}

%%%%%%%%%%%%%%%%%%%%%%%%%%%%%%%%%%%%%%%%%%%%%%%%%%%%%%%%%%%%%%%%%%%%%%%%%%%%%%%%
\section{Methodology}
\label{S:method}
%%%%%%%%%%%%%%%%%%%%%%%%%%%%%%%%%%%%%%%%%%%%%%%%%%%%%%%%%%%%%%%%%%%%%%%%%%%%%%%%
We first discuss the general workings of the IR-RON model from Daly et al.~\cite{Daly:2016}.  
Following that, we outline how the current work improves upon the existing model and create a new MON correlation.
The predictive capability of these models are presented thereafter. 
Next, we discuss the gasoline surrogate formulation framework of Ahmed et al.~\cite{Ahmed2015}, and introduce
some alterations to their methodology, including the performance parameter evaluation and additional considerations for determination of final surrogate mixtures.
Lastly, we list the neat hydrocarbon components considered for the surrogate palette.

%%%%%%%%%%%%%%%%%%%%%%%%%%%%%%%%%%%%%%%%%%%%%
\subsection{ATR-FTIR octane model}
%%%%%%%%%%%%%%%%%%%%%%%%%%%%%%%%%%%%%%%%%%%%%

Here, our work improves upon the predictive capability of the RON correlation developed by Daly et al.~\cite{Daly:2016}.
We also present a new MON correlation, created by following the same methodology as the new RON model, using PCR to correlate fuel ATR-FTIR spectra to octane ratings.
The models were further improved by extending the original training dataset to include the FACE gasolines.
Because of this addition, the performance of the model cannot be judged by how well the FACE gasolines are predicted, as originally done by Daly et al.~\cite{Daly:2016}.

Instead, model performance (foresight) will be measured by its ability to predict 25 TRF fuel mixtures from Knop et al.~\cite{Knop2014}.
We also provide a comparison of the IR model's predictive capability to seven octane correlations (specific to TRF mixtures) proposed in the literature~\cite{Knop2014}.
We performed this extra step to show that the IR-RON model of Daly et al., although not made specifically for TRF mixtures, can predict octane numbers of TRF mixtures at least as well as the TRF-specific correlations used by Sarathy et al.~\cite{Sarathy2016}.
We now outline the calculation methodology with the updated IR-octane models.

In this approach, PCR assigns a weighting coefficient to each absorbance value (for all wavenumbers), and the dot product of these weightings with a fuel's ATR-FTIR absorbance spectra, plus a constant offset factor, yields the octane rating.
In equation form, this is given by
\begin{equation}
\label{eq: octane relation}
\text{ON} = \sum_{\nu} W_{\nu}A_{\nu} + b \;,
\end{equation}
where: ON = RON or MON; $W_{\nu}$ and $A_{\nu}$ are the weighting coefficients and absorbance spectra associated with a particular wavenumber $\nu$, respectively; and $b$ is an offset. The summation is performed over the applicable wavenumber range.
The spectral weightings and offset for the RON and MON correlations can be found as supplemental material.
The absorbance spectra of a gasoline, neat hydrocarbons, or mixtures of neat hydrocarbons may be attained by using an ATR-FTIR\footnote{Here we used a ThermoFisher Nicolet iS10 FTIR with a single-bounce, Attenuated Total Reflectance (ATR) smart accessory (\SIrange{650}{4500}{\per\centi\meter^{-1}} at \SI{2}{\centi\meter^{-1}} resolution, nitrogen purged, and crystal type: diamond with \ce{ZnSe} lens, part number: 222-24700).}, with $\nu$ ranging \SIrange{650}{3580}{\centi\meter^{-1}} in increments of \SI{1}{\centi\meter^{-1}}
(See Daly et al.~\cite{Daly:2016} for more details on spectra collection methodology).
Mixtures of neat fuels are assigned an IR spectra based on the molar-weighted sum of the individual components:
\begin{equation}
A_{b} = \sum_{\nu} \sum_{j=1}^{n} X_{j}A_{j,\nu} \;,
\end{equation}
where $A_{b}$ is the resulting linearly blended spectra of mixture $b$, $X_{j}$ is the mole fraction of each component $j$, $A_{j}$ is component $j$'s spectra, and the summation is performed over the $n$ components in the mixture.
The mole-weighted summation is carried out over all wavenumbers $\nu$.
In this manner, a small database of neat hydrocarbon spectra can be used to numerically create fuel spectra.\footnote{Alcohols, such as ethanol, exhibit nonlinear IR blending due to solvation effects,~\cite{VanNess1967,Reilly2013} so precautions are needed to include alcohols. See the work of Corsetti et al.~\cite{Corsetti2015} for a relevant analysis of this phenomena.}

\begin{figure}[htbp]
\centering
\includegraphics[width=\linewidth]{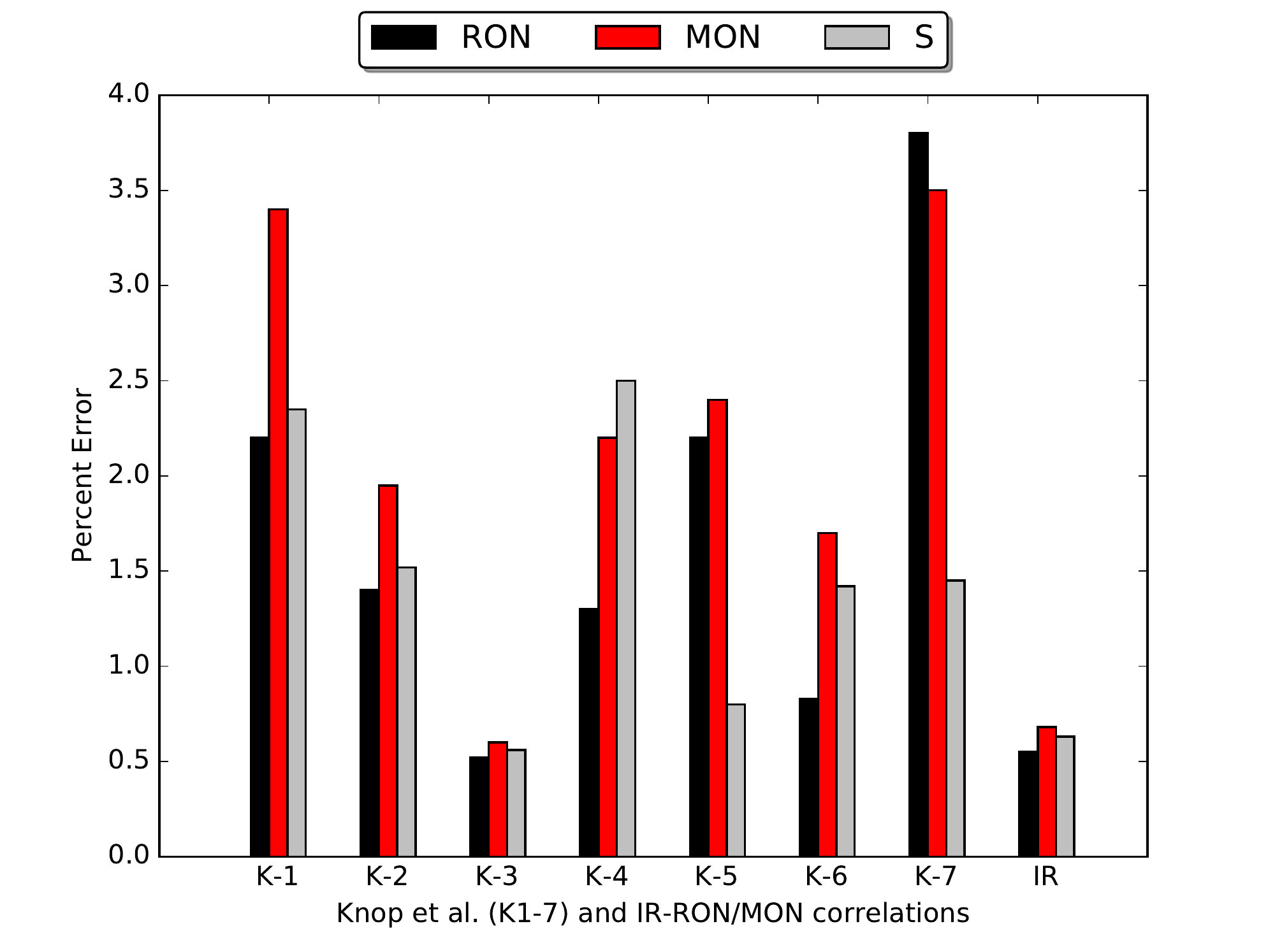}
\caption{Global error of the Knop et al.~\cite{Knop2014} and current correlations for RON, MON, and sensitivity for the TRF dataset.}
\label{F:TRF Predictions}
\end{figure}

We evaluate the newly developed models by their ability to predict the octane numbers of the TRF mixtures referenced in Knop et al.~\cite{Knop2014}.
Our updated RON and new MON correlations predicts RON, MON, and sensitivity of the TRF dataset as well as TRF model (designated here as ``K-3'') and better than the other six correlations investigated by Knop et al.~\cite{Knop2014}
Figure~\ref{F:TRF Predictions} shows the global error for RON, MON, and S for Knop et al. models (``K-1'' to ``K-7'') and our models (``IR''). 
The global error measures the average root-sum-square error of RON, MON, and sensitivity (S) over all 25 TRFs---lower values are better.

The IR correlations accurately predict quantities of TRF fuel mixtures as well or better than other proposed TRF-specific correlations in the literature~\cite{Knop2014}. 
%It also well-describe the majority of the 17 refinery-grade gasoline fuels considered, but we do not yet recommend using it for refinery-grade gasolines due to the poor prediction of a few of the 17 samples tested.  
As of now, the model is proven for TRF fuel mixtures, the 34 pure hydrocarbon components outlined in Daly et al.~\cite{Daly:2016}, fuel mixtures with various proportions of \textit{n}-heptane, isooctane, toluene, methylcyclohexane, 1-hexene, and ethanol~\cite{Foong:2014fz,Truedsson:2012iu,Perez:2012dga}, and the FACE gasolines~\cite{Cannella:2014aa}.  Based on this evidence, we believe the IR models to be valid for the hydrocarbons (and mixtures thereof) studied in this work.  

In the next section, we incorporate these RON and MON models into the surrogate formulation framework and develop surrogates for the 10 FACE gasolines.
At the time of writing this, surrogates for four of these have not yet been published by any researchers.
The surrogates are benchmarked by their ability to match the measured properties of the FACE gasolines (see Table~\ref{T:FACE_gasolines}).
We then compare formulated FACE gasoline surrogates A, C, F, G, I and J with established surrogates from the literature~\cite{Sarathy2014,Ahmed2015,Sarathy2016, Javed2017} (found in~\Cref{T:FACE_A,T:FACE_C,T:FACE_F,T:FACE_G,T:FACE_I,T:FACE_J}).

%%%%%%%%%%%%%%%%%%%%%%%%%%%%%%%%%%%%%%%%%%%%%
\subsection{Framework for formulating surrogate fuels}
%%%%%%%%%%%%%%%%%%%%%%%%%%%%%%%%%%%%%%%%%%%%%

The framework for formulating surrogate fuels builds on that of Ahmed et al.~\cite{Ahmed2015}, although our approach uses an open-source software stack based on Python.
We used Cantera~\cite{Cantera} to handle all chemical-kinetic property evaluations and as the basis for simulating autoignition delay (used for the preliminary ignition delay-octane correlation study).
Our approach does still rely on NIST's non-open REFPROP~\cite{Lemmon:2013th} software for evaluation liquid density and distillation curves.
We accessed REFPROP via Python using an established interface developed by Thelen~\cite{thelen}.
We now outline the objective function, and the methodology to calculate various performance parameters for it.

We use the objective function, the performance variables, and each variable's respective weighting factor given by Ahmed et al.~\cite{Ahmed2015}, which Table~\ref{T:Weights} shows. However, we add MON to our objective function, with the same weight as RON. Again, these are calculated based on the FTIR-octane models, as outlined in the previous subsection.     
The objective function, $f$, sums weighted errors 
in performance variables of a given surrogate with respect to the target fuel:
\begin{equation}
f = \sum_{j=1}^{n} \beta_{j}E_{j} \;,
\end{equation}
where $\beta_j$ is the weight of parameter $j$, $E_j$ is the relative error in parameter $j$ between the surrogate and the target fuel, and $n$ is the number of performance parameters (six, in this case).
Ahmed et al. chose weighting factors according to each optimization parameter sensitivity, which signifies the respective reduction of error per unit increment of the weighting factor. High sensitivity parameters, such as RON, require a larger weight to minimize surrogate error.
%See Ahmed et al. for more details~\cite{Ahmed2015}.

\begin{table}[htbp]
%\scriptsize
\centering
\begin{tabular}{@{}l c c c c c c@{}}
\toprule
 & RON & MON & Density & Distillation & C--C & H/C \\
\midrule
Weight ($\beta$) & \num{e4} & \num{e4} & \num{e3} & \num{e2} & \num{e2} & 1 \\
\bottomrule
\end{tabular}
\caption{Weights for each performance parameter considered in the objective function, taken from Ahmed et al.~\cite{Ahmed2015} (except MON, which matches RON here).
}
\label{T:Weights}
\end{table}

In addition to RON and MON, the objective function requires four additional performance parameters: hydrogen-to-carbon (H/C) ratio, distillation points (temperature at 10-80\% volume recovered), density, and carbon--carbon (C--C) bond types.
Ahmed et al.~\cite{Ahmed2015} provide more details on the calculation methodology for these parameters, though we briefly describe them here for completeness.

H/C ratio, the overall proportion of hydrogen to carbon atoms in the mixture, is calculated via the molar-weighted sum of each component species's ratio.
The distillation curves and fluid density calculations are determined in the same manner as Ahmed et al.~\cite{Ahmed2015}.
For a given fuel mixture, REFPROP approximates the mixture fluid density and determines the distillation curve by simulating thermodynamic vapor-liquid equilibrium states, described in detail by Mueller et al.~\cite{Mueller2012}
We consider neat hydrocarbons not represented in default REFPROP database; as such, they were incorporated into the software with guidance from the developers.~\cite{HuberPersonal}
This required creating reference files containing thermodynamic properties for these species, which REFPROP uses in evaluating the Peng--Robinson equation of state, as well as polynomial fits for specific heat with temperature~\cite{Lemmon:2013th}.
These hydrocarbons not included in REFPROP are summarized in Table~\ref{T:palette}, and the associated thermodynamic files are available openly~\cite{Daly2017EOS}.
Our distillation calculations produce advanced distillation curves (ADC), known to be quantitatively different from those produced from typical ASTM-D86 experimental procedures, as laid out by the works of Bruno et al.~\cite{bruno2009complex, bruno2010composition, bruno2010relating}.  Advanced distillation curves are only available for six of the ten FACE gasolines~\cite{Burger2015}.  For consistency between formulating all ten FACE surrogates, we use the complete ASTM-D86 distillation data~\cite{Cannella:2014aa} for our study.  Future efforts will address sensitivity in formulated surrogate compositions to using ADC versus ASTM-D86 data.   
Lastly, we assigned the 14 C--C bond types in the same manner as Sarathy et al.~\cite{Sarathy2016}, including their nomenclature for numbering the types:

\begin{multicols}{2}
\begin{description}[labelsep=1em, align=left, labelwidth=1em,labelindent=0em]
{\setstretch{0.25} \footnotesize
\item[1] \textit{n}-\ce{CH3} - primary carbon 
\item[2] \textit{n}-\ce{CH2} - secondary carbon
\item[3] \textit{iso}-\ce{CH} - tertiary carbon
\item[4] \textit{naphthene} - \ce{CH2} - secondary carbon
\item[5] \textit{naphthene to alkyl} - tertiary carbon
\item[6] \textit{naphthene to naphthene} - tertiary carbon
\item[7] \textit{aromatic} \ce{CH} - tertiary carbon
\item[8] \textit{aromatic to alkyl} \ce{C} - quaternary carbon
\item[9] \textit{aromatic to naphthene} \ce{C} - quaternary carbon
\item[10] \textit{aromatic to aromatic} \ce{C}- quaternary carbon
\item[11] \textit{aliphatic} \ce{C} - quaternary carbon
\item[12] \textit{primary} \ce{C=C} double bond
\item[13] \textit{secondary} \ce{C=C} double bond 
\item[14] \textit{tertiary} \ce{C=C} double bond
\item[]
}
\end{description}
\end{multicols}
%Figure~\ref{F:FGCC} shows the amounts of each C--C bond type for the FACE gasolines, based on the 161 compounds---147 specific molecules and 14 isomer groups---identified through detailed hydrocarbon analysis (DHA) by gas chromatography-flame ionization detector~\cite{Cannella:2014aa}.
%When necessary, we assumed a bond orientation for isomers, e.g., 1,2,4-trimethylcyclopentane represents ``trimethylcyclopentanes.''
%The DHA found small lumped groups of isomers---less than a summed amount of \SI{2.0}{\percent} by mole--but these may cause slight differences between other DHA results in the literature~\cite{Ahmed2015,Sarathy2016}.
%Figure~\ref{F:FGCC} shows the amounts of each C--C bond type for the FACE gasolines, based on the detailed hydrocarbon analysis (DHA) by gas chromatography-flame ionization detector~\cite{Cannella:2014aa}.
Table~\ref{T:FACE_gasolines} lists the amounts of each C--C bond type for the FACE gasolines, based on the detailed hydrocarbon analysis (DHA) by gas chromatography-flame ionization detector~\cite{Cannella:2014aa}.
The general gasoline formulation framework approach will now be explained.  

The surrogate formulation framework's task, given a user-defined species palette, is to determine a set of optimal mole fractions that minimize the objective function; i.e., numerically blend species to closely match the fuel properties of the target fuel.  This requires iteration, where each iteration is a new set of mole fractions that dictate the bulk fuel properties.  For example, in the case of calculating RON and MON, linear-blended spectra (by mole) based on each species in the palette is first determined; then, the spectra is fed to the FTIR-octane models to calculate RON and MON.  The errors between the predicted RON and MON and the target fuel RON and MON contribute to the overall objective function value. 
To find the global minimum of the objective function, we used a constrained multiobjective optimization routine from the SciPy library (\texttt{scipy.optimize.minimize()})~\cite{SciPy}, with a constrained quasi-Newton method (L-BFGS-B)~\cite{zhu1997algorithm}.
We also found, in agreement with Ahmed et al.~\cite{Ahmed2015}, that the optimization algorithm could get ``stuck'' in many local minima of the objective function. 
To resolve this, we fed 100 randomized initial conditions (i.e., mixture mole fractions) into the optimization routine, then chose as the final result the designed surrogate (out of 100) with the lowest objective function value.
We acknowledge that other optimization routines may be more efficient than this ad-hoc solution, such as a basin-hopping routine (e.g., \texttt{scipy.optimize.basinhopping()}), but our approach requires minimal computational effort to redundantly evaluate the objective function, so we did not see much reason to pursue alternatives.

In the optimization algorithm, we constrain the mole fractions for each constituent species in the surrogate palette between zero and a species-dependent value less than one.
The upper bound of each species is dictated by its respective hydrocarbon class and the relative proportion of that class within the target fuel, as per the DHA.
For example, if the target fuel has \SI{20}{\percent} by mole of \textit{n}-paraffins, and the surrogate palette has two \textit{n}-paraffins such as \textit{n}-pentane and \textit{n}-heptane, then the upper mole-fraction bound of both species is 0.2.
This seemed a reasonable bound to expedite the surrogate design as opposed to setting the bound to one.
However, if the algorithm designs a final surrogate with a species at its upper bound, then the bound is increased and the optimization repeated to (potentially) yield a better solution.
We also ensured that species mole fractions summed to one by applying a penalty function to the objective (multiplying by \num{e8}) when violating this constraint.

We also imposed additional logic at the end of the optimization routine to further simplify surrogates: if the number of species in the palette is greater than seven, and if any these species are present at less than \SI{4}{\percent} by mole, then a new surrogate palette is generated without these minor species. This threshold of \SI{4}{\percent} was arbitrary and deemed to quantify a ``minor'' amount.  We recognize that the threshold method could possibly cause any given hydrocarbon class containing only ``minor'' species to be completely eliminated for the reduced palette.  We did find that this does occur, but with no penalty to the objective function score for the ``reduced'' formulated surrogate.
We then re-execute the surrogate formulation framework with this reduced palette, and again with 100 new randomized fuel mixture initial conditions.
We repeat this process until no minor species remain in the final surrogate---we present this result as a ``reduced'' surrogate, versus the ``full'' surrogate before this process is applied.
This was performed in anticipation of future efforts that will require reduced chemical kinetic models for these FACE surrogates; the kinetic model of Sarathy et al.~\cite{Sarathy2016}, which incorporates 2315 species and 10,079 reactions, is extensive---fewer components in the final surrogate mixture will result in a simpler reduced kinetic model.

Table~\ref{T:palette} shows the palette species chosen.
We chose the species palette constrained to three criteria.  First, we utilize hydrocarbons included in prior surrogate research~\cite{Mehl2011NTC,Sarathy2014,Ahmed2015,Sarathy2016, Javed2017}.
Next, we only considered species available in comprehensive chemical kinetic models, so that surrogates may be readily used for computational studies; the comprehensive model of Sarathy et al.~\cite{Sarathy2016} includes a wide array of species.  This kinetic model has additional species not included in past surrogate research.
Lastly, we selected species present in the training dataset of the IR-octane correlation used in this work; i.e., the species that informed the PCR model and whose properties we expect to be adequately predicted.
See Daly et al.~\cite{Daly:2016} for the complete list of fuels used in the training data set.
One exception to this is pure \textit{n}-pentane, which is not present in the training dataset directly as a pure component for the IR-octane models.
The FACE gasolines themselves are included in the training dataset in this work, and contain \SIrange{0.06}{15.06}{\percent} by mole \textit{n}-pentane as per the DHA.
As such, we expect fuels with less than \SI{15}{\percent} by mole of \textit{n}-pentane to be adequately predicted; prediction error may increase for fuels with more than \SI{15}{\percent} of \textit{n}-pentane. The pure component spectra of \textit{n}-pentane was collected after the IR-RON and MON model training phase, and will be included in future efforts.  \textit{N}-butane was not able to be included in the species palette, as with literature surrogates~\cite{Ahmed2015,Sarathy2016}, since we could not procure the ATR-FTIR spectra.  This is primarily due to limitations in the spectra collection methodology, which require the substance measured to be solid or liquid at room conditions. 

Based on our constraints, the final fuel palette extends previous efforts by including n-pentane, 2-methylpentane, 1-pentene, cyclohexane, and o-xylene.  These species broaden the range of molecular weights and hydrogen to carbon ratios for each PIONA class, present to varying extents within the FACE gasolines.  Indeed, many of these species are not prevalent in most of the FACE gasolines individually.  However, guided by the work of Daly et al.~\cite{Daly:2016}, blended octane ratings are highly sensitive and non-linear to additions of these species.  As such, incorporating these species extends the range of blended octane rating combinations, and we expect better convergence to target octane ratings.  We limit the parameterization of our current study by omitting some candidate species, such as m-/p-xylenes or 2-pentene; these will be investigated in future efforts.  Although, as we will show, there are marginal benefits to leveraging an expansive species palette as opposed to a reduced sub-set.      

\begin{table}[htbp]
\centering
\begin{tabular}{@{}l c c c c c c@{}}
\toprule
Palette species & CAS \# & Formula & Class & $T_{b}$ & MW & REFPROP\\
&&& (PIONA) & [\si{\kelvin}] & [\si{\gram \per \mole}]\\
\midrule
\textit{n}-pentane	&  109-66-0 &\ce{nC5H12}     & paraffin & 309.2 & 72.14 & $\checkmark$\\
\textit{n}-heptane	& 142-82-5 &\ce{nC7H16}     & paraffin & 371.5 &100.20 & $\checkmark$ \\
2-methylbutane	     & 78-78-4 &\ce{C5H12}     & isoparaffin & 300.9 &72.14 & $\checkmark$\\
2-methylpentane		 & 107-83-5  & \ce{C6H14} 		& isoparaffin & 333.3& 86.17 & $\checkmark$ \\
2-methylhexane	     &  591-76-4 &\ce{C7H16}   	& isoparaffin & 363.4  & 100.21 & $\times$ \\
2,2,4-trimethylpentane	& 540-84-1  & \ce{C8H18}    	& isoparaffin & 372.3 & 114.22 & $\checkmark$\\
1-pentene	    & 109-67-1 &\ce{C5H10}    	& olefin  & 304.0 & 70.13 & $\times$\\
1-hexene	    & 592-41-6 &\ce{C6H12}     	& olefin & 337.0 & 84.15 & $\times$\\
cyclopentane	& 287-92-3 &\ce{C5H10}	 	& naphthene & 322.0 & 70.13 & $\checkmark$\\
cyclohexane     & 110-82-7 &\ce{C6H12}     & naphthene & 353.8 & 84.15 & $\checkmark$\\
toluene	        & 108-88-3 &\ce{C7H8}   	& aromatic & 383.7 &92.13& $\checkmark$\\
\textit{o}-xylene		        & 95-47-6 &\ce{C8H10}	 & aromatic & 417.5& 106.16 &$\checkmark$\\
1,2,4-trimethylbenzene	    & 95-63-6 &\ce{C9H12}    & aromatic & 442.4 & 120.19 & $\times$ \\
1,3,5-trimethylbenzene	    & 109-67-1 &\ce{C9H12}    & aromatic & 437.8 & 120.19 & $\times$ \\
\bottomrule
\end{tabular}
\caption{
Species palette for FACE surrogates. $T_{b}$ is the normal boiling point, and ``REFPROP'' indicates whether the species was represented in the software. Species not included in the default REFPROP database were manually added. 
}
\label{T:palette}
\end{table}
\clearpage

%%%%%%%%%%%%%%%%%%%%%%%%%%%%%%%%%%%%%%%%%%%%%%%%%%%%%%%%%%%%%%%%%%%%%%%%%%%%%%%%%%%%%%%%%%%%%%%%%%%%
\section{Results and discussion}
\label{S:results}
%%%%%%%%%%%%%%%%%%%%%%%%%%%%%%%%%%%%%%%%%%%%%%%%%%%%%%%%%%%%%%%%%%%%%%%%%%%%%%%%%%%%%%%%%%%%%%%%%%%%

First, we present the surrogates developed here for FACE gasolines A--J.
These surrogates were generated from the full, original species palette with 8--13 species (depending on the target FACE gasoline).
We also developed simpler surrogates containing 4--7 species from a reduced species palette.
The supplemental material contains tables and figures comparing the full- and reduced-palette surrogates, describing in detail how well the surrogate matches respective target properties.
Finally, we compare the surrogates for FACE gasolines A, C, F, G, I and J with other proposed surrogates from the literature.

%%%%%%%%%%%%%%%%%%%%%%%%%%%%%%%%%%%%%%%%%%
\subsection{Full-palette FACE surrogates}

Table~\ref{T:FACE_Surr_Full} presents the full-palette FACE gasoline surrogates.
A molar amount of zero indicates that the species was considered but the algorithm converged on that value, while a blank entry indicates the species was not considered.

The FACE gasoline surrogates, on average, match the array of target properties within \SI{5}{\percent}.
This demonstrates that our framework can create gasoline surrogates for the wide range of gasoline target properties the FACE gasolines represent---given the species palette utilized.
In contrast to the other surrogates generated, the FACE G surrogate exhibits high prediction errors in density (\SI{5.3}{\percent}) and H/C ratio (\SI{16.8}{\percent}).
This may result from the relatively high weighting factors used to (successfully) match RON and MON with low errors.
At the same time, the species mole fractions were not strictly constrained to enforce matching the target H/C, \ce{C-C} types, or density; as a result, the algorithm was free to depart from these target properties to minimize the objective function.
To match the high octane sensitivity of FACE G---11, the highest of the FACE gasolines---the optimization algorithm selected large amounts of olefins (high sensitivity) and low amounts of \textit{n}-paraffins (low sensitivity); no other options were available based on the species present in the palette.
%We conclude that the species palette, and the IR-octane models, could be extended with additional high-sensitivity fuel mixtures for greater model robustness, as it should be possible to design a high sensitivity fuel with proper \textit{n}-paraffin, isoparaffin, olefin, naphthene, and aromatic proportions.
We conclude that the the IR-octane models should be updated with additional high-sensitivity fuel mixtures for greater model robustness, as it should be possible to design a high sensitivity fuel with proper n-paraffin, isoparaffin, olefin, naphthene, and aromatic proportions.  This means the IR-octane model training procedure revisited by including additional high-sensitivity fuel mixtures, potentially adding new components to the surrogate species palette that the IR-octane model is updated with, followed by repeating the surrogate formulation.  
Surrogates for the remaining, lower-sensitivity FACE gasolines (ranging 0--8) adequately capture the target properties; for example, the FACE D surrogate matches all performance attributes within \SI{3}{\percent}.

\begin{table}[htbp]
\centering
\footnotesize
\begin{tabular}{@{}l c c c c c c c c c c@{}}
\toprule
% &\multicolumn{10}{c}{Full palette FACE surrogates} \\
% %
% \cmidrule{2-12}
Parameter 	& A & B & C & D & E & F & G & H & I & J \\
\midrule
RON 	        &83.9 & 95.1 &84.1	 & 93.7 & 87.4 & 93.6 & 95.9 &	86.7 &	70.2  & 72.7 \\
MON		        &83.6 & 93.2 &83.2	 & 87.7 & 81.2 & 88.2 & 86.7 &	80.1 &	69.5  & 71.0 \\
Density (\si{\kilo\gram\per\meter^3})			&697  & 709  &703    & 762  & 734  & 734  & 803  &	771  &  716   & 748  \\
H/C				&2.23 & 2.15 &2.16   & 1.83 & 2.00 & 2.03 & 1.57 &	1.76 &	2.12  & 1.90 \\
\midrule
\% Volume distilled	& \multicolumn{10}{c}{Temperature [\si{\kelvin}]}  \\
\midrule
10			& 338 & 339 &   336	& 347 &337 & 345& 348& 344&343 &352 \\
20		    & 344 & 346 &   341	& 354 &342 & 350& 358& 349&348 &360 \\
30	        & 352 & 355 &   348 & 361 &346 & 356& 372& 355&354 &368 \\
40	        & 360 & 365 &	355	& 370 &352 & 363& 389& 364&360 &375 \\
50	        & 367 & 372 &	363	& 378 &358 & 369& 406& 374&367 &381 \\
60	        & 370 & 375 &	370	& 387 &365 & 374& 418& 387&370 &388 \\
70	        & 371 & 377 &	374	& 395 &371 & 380& 423& 399&373 &400 \\
80	        & 372 & 380 &	378	& 405 &379 & 387& 426& 413&378 &422 \\
\midrule
Carbon type	& \multicolumn{10}{c}{Fractional \%}  \\
\midrule
1&	54.8&	55.4&	51.8&	38.9&	33.3&	42.5&	24.1&	27.5&	41.4&	33.6\\
2&	21.5&	14.9&	21.5&	14.3&	18.9&	15&	14.8&	17.4&	27.9&	25.8\\
3&	11.9&	12.1	&11.4&	7.4	&6.7&	8.5&	0.3&	2.8&	7&	3.1\\
4&	0&	0&	0&	3.1&	19.4&	11.9	&6.6&	14.2&	7.9	&7.2\\
5&	0&	0&	0&	0&	0&	0&	0&	0&	0&	0\\
6&	0&	0&	0&	0&	0&	0&	0&	0&	0&	0\\
7&	3.4&	6.9&	6.7&	23.6&	9.4&	9.5&	28.1&	23.3&	7.3&	13.2\\
8&	1&	3.1&	2.8&	10&	4&	4.7&	17.1&	10.8&	3.3	&9.9\\
9&	0&	0&	0&	0&	0&	0&	0&	0&	0&	0\\
10&	0&	0&	0&	0&	0&	0&	0&	0&	0&	0\\
11&	7.4&	7.7&	5.8&	2.7&	3.1&	5.6&	0.3&	1.2&	4.2	&2.5\\
12&	0&	0&	0&	0&	2.6&	1.2&	4.4	&1.3	&0.5	&2.4\\
13&	0&	0&	0&	0&	2.6&	1.2	&4.4&	1.3	&0.5&	2.4\\
14&	0&	0&	0&	0&	0&	0&	0&	0&	0&	0\\

\midrule
Species & \multicolumn{10}{c}{Molar \%} \\
\midrule
\textit{n}-pentane 	    &   6.4    & 0.0  & 5.9      &0.3      &      &    0.0   &          & 5.3    & 13.0 & 3.0  \\
\textit{n}-heptane	    &   5.1    & 1.7  & 5.3      &4.1      & 1.3  &          &    3.9   & 9.4    & 15.0 & 21.8 \\
2-methylbutane	        &   20.0   & 24.0 & 18.6     &9.8      & 10.2 &    9.4   &    0.0   &        & 7.7  & 2.3  \\
2-methylpentane		    &   1.0    & 4.2  & 13.3     &20.4     & 1.0  &    0.1   &    0.5   & 7.8    & 4.2  & 1.9  \\
2-methylhexane	        &   10.6   & 3.3  & 6.5      &3.1      & 12.6 &    10.4  &          & 2.8    & 7.0  &      \\
2,2,4-trimethylpentane	&   51.8   & 54.9 & 40.0     &18.9     & 19.9 &    38.9  &    1.8   & 8.3    & 28.5 & 18.1 \\
1-pentene	            &          &      &          &         & 4.8  &          &    11.2  & 5.0    & 0.6  & 8.4  \\
1-hexene	            &          &      &          &         & 12.1 &    8.3   &    20.1  & 4.0    & 3.0  & 8.6  \\
cyclopentane	        &          &      &          & 4.3     & 15.6 &    16.5  &    9.3   & 18.6   & 0.0  &      \\
cyclohexane             &          &      &          &         & 8.0  &          &          & 0.5    & 9.0  & 8.5  \\
toluene	                &  3.4     & 1.9  & 4.3      & 8.8     & 4.1  &          &          & 13.6   & 5.1  & 3.4  \\
\textit{o}-xylene		    &  1.7     & 9.9  & 5.0      & 30.1    & 9.6  &    16.4  &    38.9  & 14.9   & 3.3  & 5.0  \\
1,2,4-trimethylbenzene	&          &      & 1.5      & 0.3     &      &          &    14.4  & 9.8    & 3.7  & 15 \\
1,3,5-trimethylbenzene	&          &      &          &         & 0.8  &          &          &        &      &      \\
\midrule
Objective Function & 1.61 & 2.85 & 1.25 & 3.12 & 0.30 & 1.91 & 10.5 & 3.29 & 3.05 & 6.95\\
\bottomrule
\end{tabular}
\caption{
Full palette, formulated FACE gasoline surrogates.  A blank entry indicates that the species was considered in the work, but was not included in the palette for the particular FACE gasoline. A zero (0) indicates the species was in the palette, but not chosen by the optimizer.
}
\label{T:FACE_Surr_Full}
\end{table}

\subsection{Reduced-palette FACE surrogates}

Table~\ref{T:FACE_Surr_Reduced} presents the simpler FACE surrogates, with four to seven species, designed using reduced species palettes.
Interestingly, though these surrogates contain fewer components, they capture the properties of the target fuels better, or nearly as well as the larger, full-palette surrogates.  The minor species removed from the ``full'' palette did result in some hydrocarbon classes being completely eliminated.  For example ``reduced'' surrogates for FACE A and B no longer have \emph{n}-paraffins, and the objective function score was further minimized (indicating a better surrogate).
The reduced-palette surrogates, on average, predict RON, MON, and distillation curve slightly better, and density and H/C ratio slightly worse than the full-palette surrogates.  Overall, the reduced-palette FACE surrogates achieved better objective function evaluations than their full-palette counterparts---meaning they better-match the real gasoline.  
Based on the overall similarity in performance between the two sets of surrogates, satisfactory surrogate fuels may be obtained without requiring an ever-larger palette of potential components. 
This result is counter intuitive. We expect that as components in the surrogate palette and their relative proportions approaches that of the target fuel the objective function would be best-minimized (approach zero).
We suggest that modeling artifacts causes much of the ``reduced'' surrogates to outperform the ``full''. The overall accuracy of predicted fuel properties could be reducing as the species palettes grows.  Larger species palettes also bring the increased possibility that the optimization routine is not guaranteed to return a global minimum. Additionally, weighting factors for the objective function may also need to be tuned on a per-surrogate basis, due to the large variability in the species palette and therefore parameter sensitivity.  These intricacies and their impact on modeling results were not investigated in this work, but should be investigated in future efforts.  

Unfortunately, limiting the palette exacerbates some errors for the worse-performing full-palette surrogates.
For example, FACE G reduced-palette surrogate captures H/C ratio worse at \SI{15.7}{\percent}, although density, RON, and MON still closely match those of the full-palette surrogate.
The supplemental material contains surrogate-specific target property changes between the full- and reduced-palette formulations.
In general, the \ce{C-C} bond types are the hardest properties to predict with the reduced palette.

\begin{table}[htbp]
\centering
\footnotesize
\begin{tabular}{@{}l c c c c c c c c c c @{}}
\toprule
% &\multicolumn{10}{c}{\small{Reduced palette FACE surrogates}} \\
% %
% \cmidrule{2-12}
Parameter 	& A & B & C & D & E & F & G & H & I & J \\
\midrule
RON 	        &83.8 & 95.1 &84.4	 & 93.9 & 87.4 & 94.4 & 96.3 &	86.9 &	70.1  & 73.2 \\
MON		        &83.7 & 93.1 &82.9	 & 87.3 & 81.1 & 88.9 & 86.8 &	79.9 &	69.5  & 70.2 \\
Density (\si{\kilo\gram\per\meter^3})			&700  & 718  &698    & 761  & 731  & 732  & 802  &	775  &  710   & 761  \\
H/C				&2.22 & 2.11 &2.20   & 1.83 & 2.05 & 2.03 & 1.54 &	1.75 &	2.15  & 1.86 \\
\midrule
\% Volume distilled	& \multicolumn{10}{c}{T [\si{\kelvin}]}  \\
\midrule
10			& 337 & 341 &   334	& 344 &339 & 346& 352& 348&341 &360 \\
20		    & 343 & 349 &   340	& 352 &342 & 351& 363& 354&347 &368 \\
30	        & 352 & 360 &   348 & 362 &346 & 357& 374& 362&353 &374 \\
40	        & 361 & 371 &	358	& 373 &351 & 363& 387& 372&360 &379 \\
50	        & 368 & 376 &	367	& 382 &357 & 369& 404& 382&364 &384 \\
60	        & 371 & 378 &	372	& 389 &364 & 374& 419& 391&367 &391 \\
70	        & 372 & 380 &	375	& 395 &372 & 379& 424& 400&368 &401 \\
80	        & 375 & 385 &	378	& 404 &382 & 386& 426& 417&369 &419 \\
\midrule
Carbon type	& \multicolumn{10}{c}{Fractional \%}  \\
\midrule
1&	54.4&	54&	53.7&	38.6&	35.4&	43.1&	23.7&	27&	44&	30.8\\
2&	21&	14.3&	21.7&	13.9&	18.4&	15.5&	17.7&	16.4&	28.4&	24.6\\
3&	13.5&	11.8&	11.5&	6.6&	7.8&	8.4&	0&	2&	9.1&	2.3\\
4&	0&	0&	0&	4.9&	20.9&	10.3&	0&	16.6&	4.2	&10.2\\
5&		0&	0&	0&	0&	0&	0&	0&	0&	0&	0\\
6&		0&	0&	0&	0&	0&	0&	0&	0&	0&	0\\
7&	3.4&	8.4&	4.3&	23&	7.3&	9&	28.6&	22.8&	8.4&	14.6\\
8&	1.7	&4.2&	2.1&	9.8&	3.6&	4.5&	17.4&	10.9&	1.7&	10.5\\
9&	0&	0&	0&	0&	0&	0&	0&	0&	0&	0\\
10&	0&	0&	0&	0&	0&	0&	0&	0&	0&	0\\
11&	6&	7.3&	6.6&	3.2&	3.5&	5.9&	0&	2&	4.2&	2.3\\
12&	0&	0&	0&	0&	1.6&	1.7&	6.3&	1.2&	0&	2.4\\
13&	0&	0&	0&	0&	1.6&	1.7&	6.3&	1.2&	0&	2.4\\
14&	0&	0&	0&	0&	0&	0&	0&	0&	0&	0\\

\midrule
Species palette & \multicolumn{10}{c}{Molar \%} \\
\midrule
\textit{n}-pentane 	    &          &      &          &         &      &          &          &        & 11.0 &      \\
\textit{n}-heptane	    &          &      & 13.0     & 8.6     &      &          &          & 16.5   & 11.6 & 23.7 \\
2-methylbutane	        &   26.1   & 23.3 & 28.2     &16.2     & 8.8  &    9.4   &          &        & 12.8 &      \\
2-methylpentane		    &          &      & 5.5      &7.7      &      &          &          &        &      &      \\
2-methylhexane	        &   25.9   & 9.1  &          &0.0      & 19.2 &     7.5  &          &        & 20.0 &      \\
2,2,4-trimethylpentane	&  42.1    & 52.4 & 45.9     &22.5     & 22.5 &    41.5  &          & 13.9   & 28.5 & 16.4 \\
1-pentene	            &          &      &          &         &      &          &9.5       & 8.0    &      & 8.5  \\
1-hexene	            &          &      &          &         & 10.3 &    11.6  &35.7      &        &      & 8.7  \\
cyclopentane	        &          &      &          & 6.9     & 27.3 &    14.3  &          & 22.8   &      &      \\
cyclohexane             &          &      &          &         &      &          &          &        & 4.7  & 12.3 \\
toluene	                &          &      &          & 8.1     &      &          &          & 15.0   & 11.4 &      \\
\textit{o}-xylene		&  5.9    & 15.2  & 7.4      & 30.0    & 11.9 &    15.7  &40.0      & 11.1   &      & 15.0 \\
1,2,4-trimethylbenzene	&          &      &          &         &      &          &14.8      & 12.7   &      & 15.4 \\
\midrule 
Objective Function & 1.21 & 2.11 & 0.60 & 2.19 & 0.48 & 0.66 & 12.2 & 2.98 & 1.87 & 4.02\\
\bottomrule
\end{tabular}
\caption{
Reduced palette FACE gasoline surrogates. A blank entry indicates that the species was considered, but was not included in the palette for the particular FACE gasoline. A zero (0) indicates the species was in the palette, but not chosen by the optimizer.  The reduced species palette was generated using an iterative, auto-reduction strategy based on a low-amount species threshold.
}
\label{T:FACE_Surr_Reduced}
\end{table}

%%%%%%%%%%%%%%%%%%%%%%%%%%%%%%%%%%%%%%%%%%%%%%%%%%
\subsection{Comparison with literature surrogates}

This section compares the surrogates developed in the current study with surrogates proposed in the literature for FACE gasolines A~\cite{Sarathy2014,Ahmed2015}, C~\cite{Sarathy2014,Ahmed2015}, F~\cite{Sarathy2016}, G~\cite{Sarathy2016}, I~\cite{Javed2017} and J~\cite{Javed2017}.  For now, we omit comparing to the FACE A, C, I and J primary reference fuel (PRF) surrogates from Shankar et al.~\cite{shankar2016primary}, as they only considered RON and MON in their formulation methodology.  While these PRF surrogates are well-validated to emulate combustion behavior for these FACE gasolines at pre-mixed conditions, the surrogates may not be applicable to more complex physical environments including spray, mixing and diffusion.  As such, we only consider the literature surrogates that targeted both physical and chemical properties, specifically: hydrocarbon classes, \ce{C-C} bond type proportions, distillation characteristics, H/C ratios, density, and octane ratings.
~\cref{T:FACE_A,T:FACE_C,T:FACE_F,T:FACE_G,T:FACE_I,T:FACE_J} compare the new surrogates with previous surrogates in terms of target properties.

\subsubsection{FACE A}

\begin{table}[htbp]
\centering
\footnotesize
\begin{tabular}{@{}lcccccc@{}}
\toprule
%& & \multicolumn{5}{c}{Surrogates}\\
%\cmidrule{3-7}\cmidrule{9-12}
%
Parameter 	&	FACE A\cite{Cannella:2014aa} & Full & Reduced & ~~Sarathy\cite{Sarathy2014} & \multicolumn{2}{c}{~~Ahmed\cite{Ahmed2015}} \\
\midrule
RON 	&83.9 & 83.9 & 83.8	 &	84.0 &86.6 &85.6 \\
MON		&83.5 & 83.6  &	83.6 &  84.0 & & \\
Density	[\si{\kilo\gram\per\meter^3}) &685 & 697 &	699.8	 &   686  & 694 & 691 \\
H/C		&2.29 & 2.22 &	2.22 &	2.28  &2.28&2.26 \\
\midrule
\% Volume distilled	& \multicolumn{6}{c}{Temperature [\si{\kelvin}]} \\
%\cmidrule{1}\cmidrule{2-7}
\midrule
10			&329 & 337 & 336 & 335 &321& 351\\
20		    &344 & 343 & 343 & 347 &337& 360\\
30	        &357 & 352 & 352 & 356 &351& 365\\
40	        &365 & 360 & 361 & 362 &361& 368\\
50	        &368 & 367 & 368 & 366 &366& 369\\
60	        &370 & 370 & 371 & 368 &368& 370\\
70	        &372 & 371 & 372 & 369 &369& 370\\
80	        &374 & 372 & 375 & 370 &369& 370\\
\midrule
Species  & \multicolumn{6}{c}{Molar \%} \\
\midrule
\textit{n}-butane       &&      &     & 7.0 & 7.7 & 5.0\\
\textit{n}-pentane	    && 6.4  & 0   & & & \\
\textit{n}-heptane	    && 5.1  & 0   & 7.0 & 10.0 & 5.0\\
2-methylbutane	        && 20.0 &26.1 & 15.0 & 12.0 & 5.0\\
2-methylpentane		    && 1.0  & 0   & & & \\
2-methylhexane	        && 10.6 & 25.9& 11.0&10.3&15.0\\
2,2,4-trimethylpentane	&& 51.8 & 42.1& 60.0&60.0&70.0\\
toluene	                && 3.4  & 0   & 0 & 0 & 0 \\
\textit{o}-xylene       && 1.7  & 5.9 & & & \\
1,2,4-trimethylbenzene	& & 0 & 0 & & & \\
\bottomrule
\end{tabular}
\caption{
The full- and reduced-palette FACE gasoline A surrogates compared with literature surrogates and the real FACE A properties. 
A blank entry indicates the species\slash parameter was not considered. A zero (0) indicates the species was in the palette, but not chosen by the optimizer.
}
\label{T:FACE_A}
\end{table}

\begin{figure}[htbp]
\centering
\begin{subfigure}[b]{.5\linewidth}
\includegraphics[width=85mm]{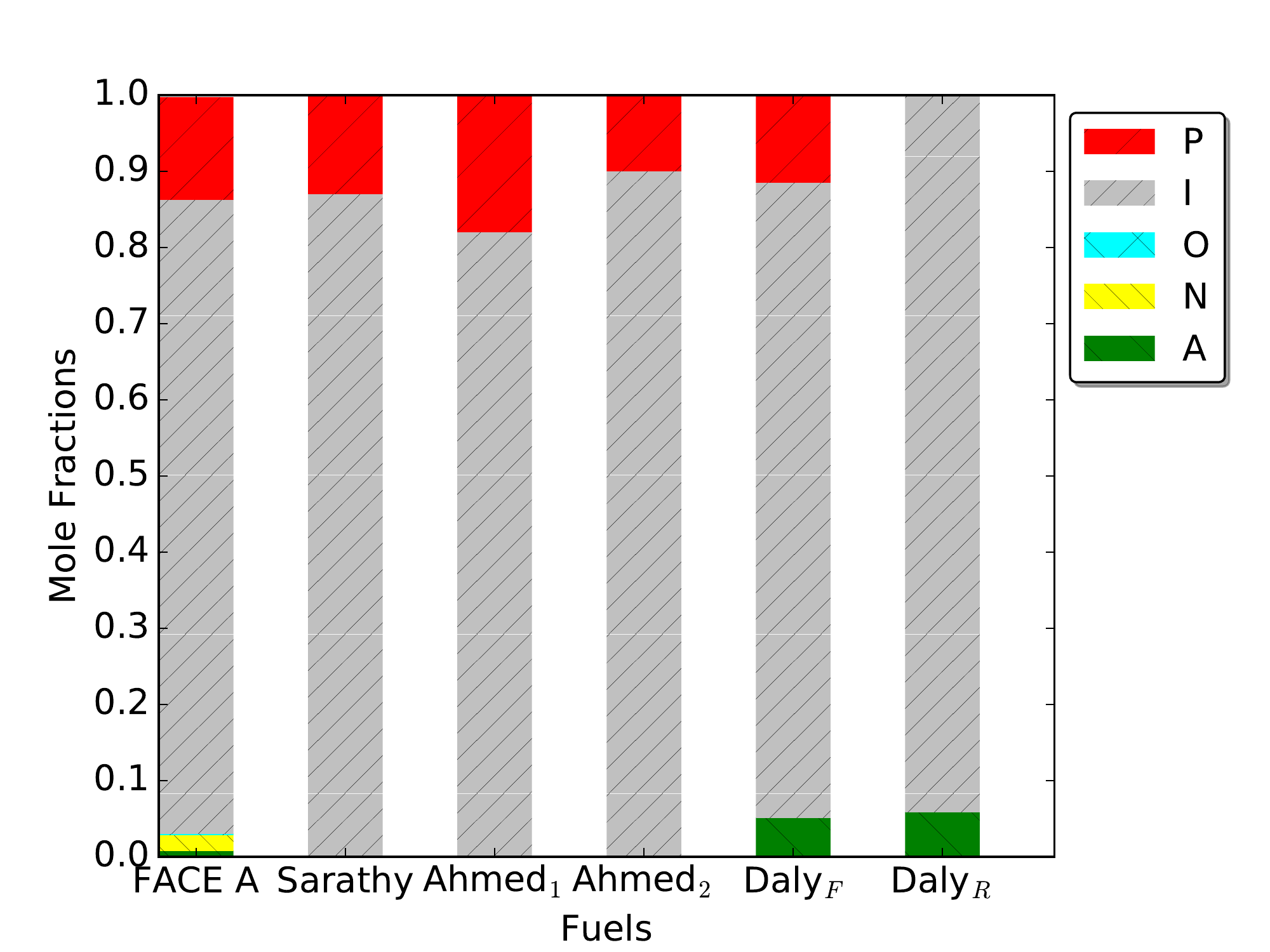}
\caption{Hydrocarbon class proportions}
\label{F:FGApiona comparison}
\end{subfigure}
~
\begin{subfigure}[b]{.45\linewidth}
\includegraphics[width=85mm]{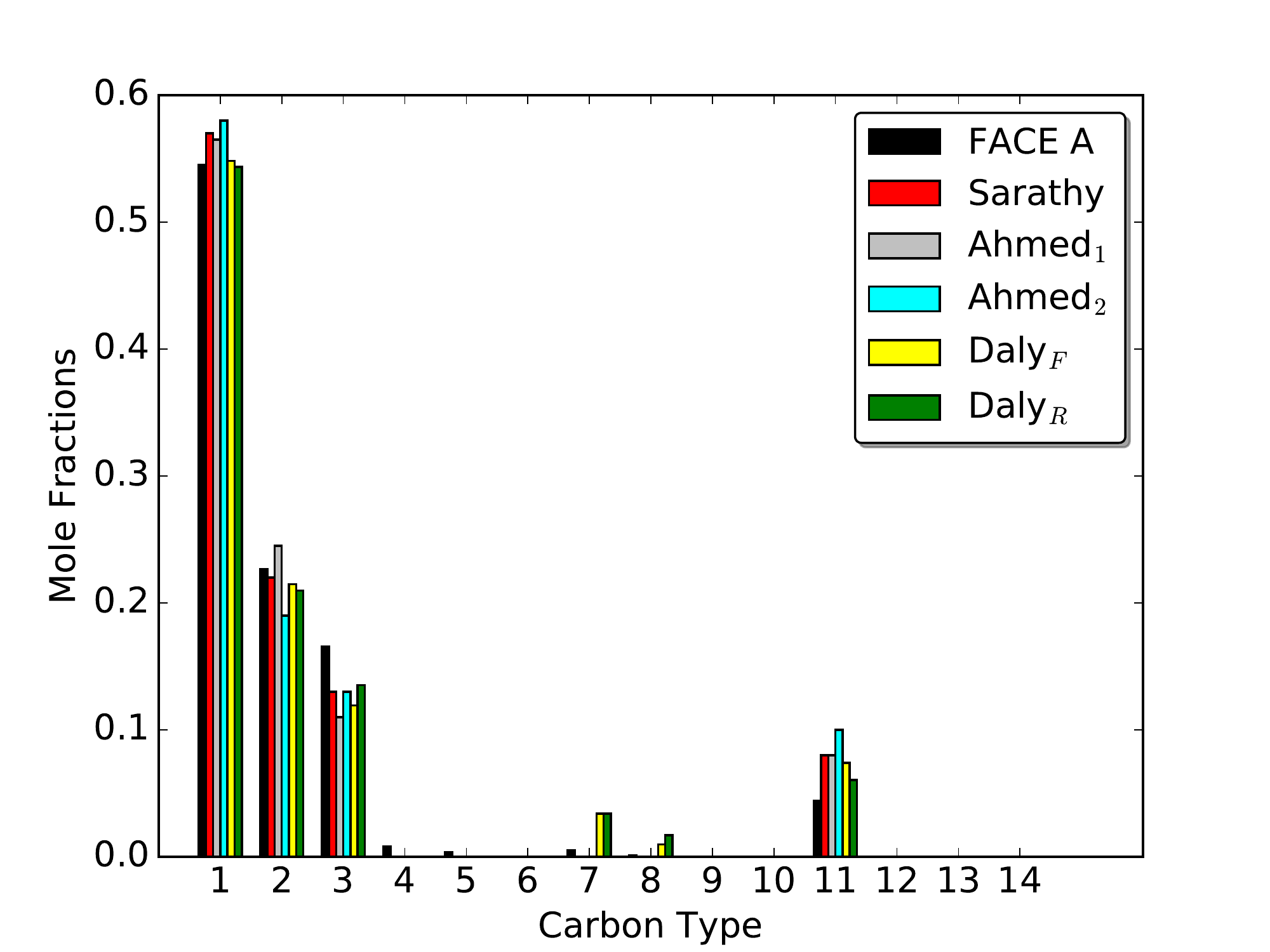}
\caption{C--C bond type proportions}\label{F:FGAcarbon comparison}
\end{subfigure}
\\
\begin{subfigure}[b]{.5\linewidth}
\includegraphics[width=85mm]{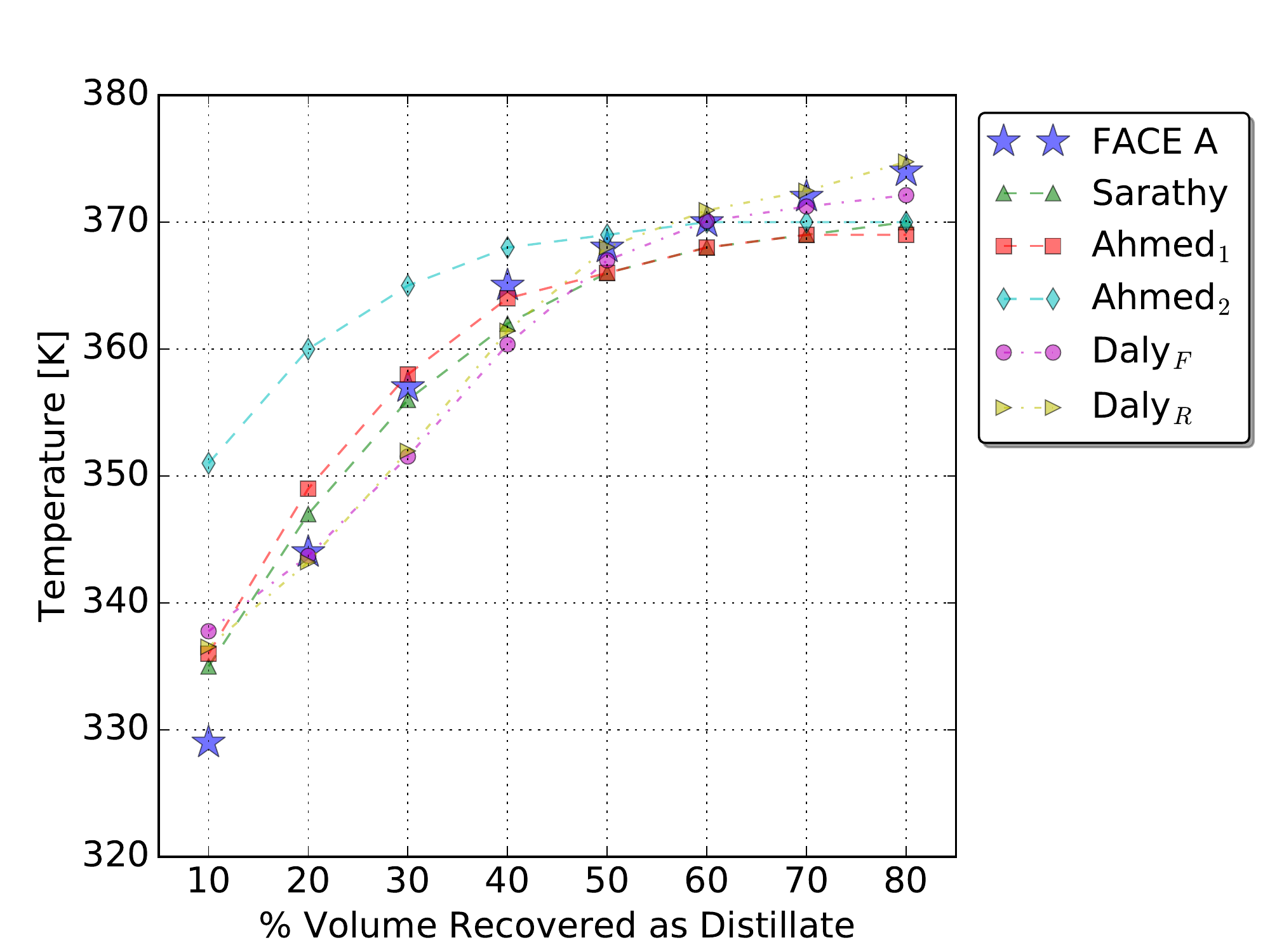}
\caption{Distillation curve}
\label{F:FGAdist comparison}
\end{subfigure}
\begin{subfigure}[b]{.45\linewidth}
\includegraphics[width=85mm]{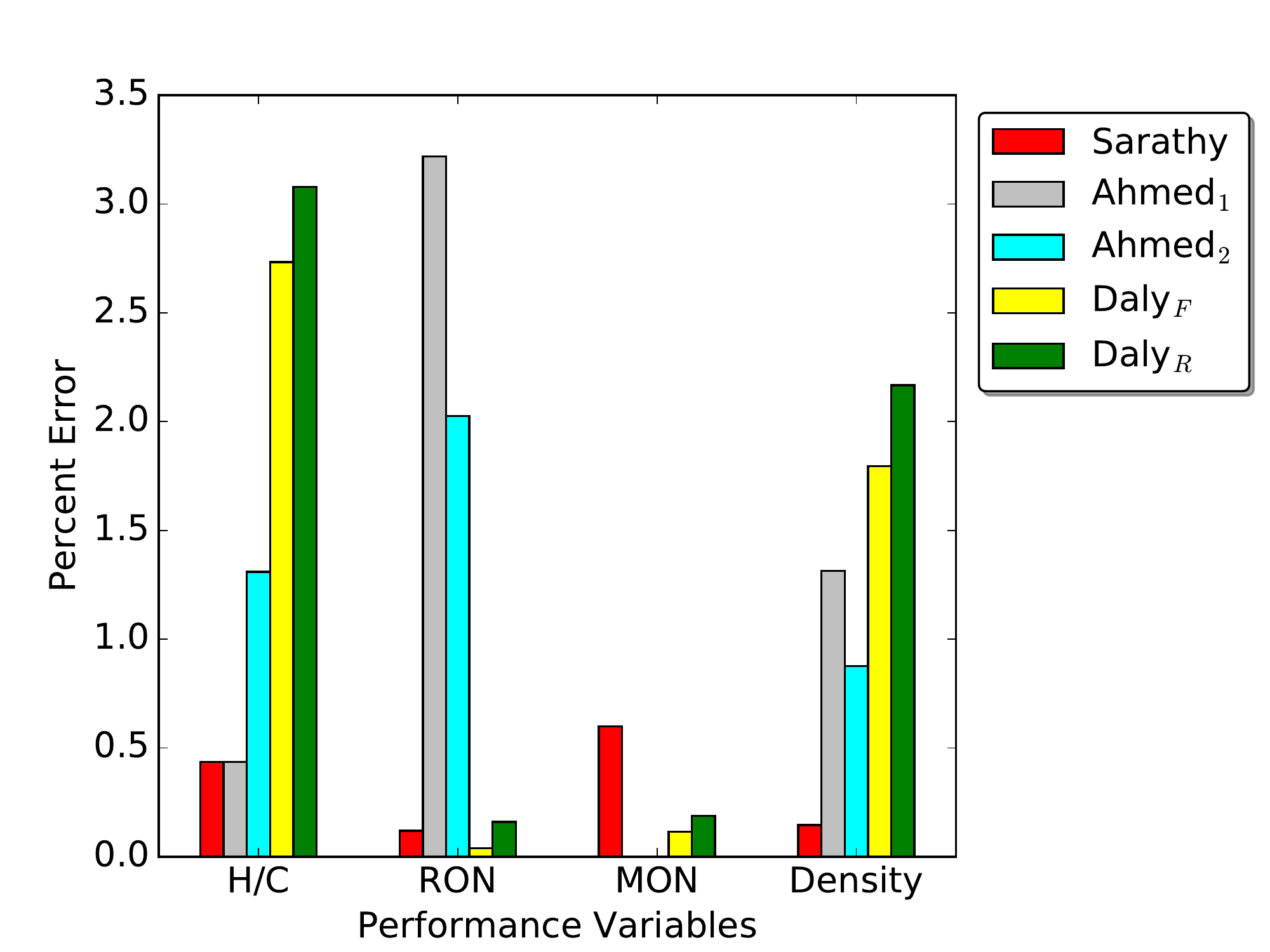}
\caption{Error in H/C, RON, MON, and density}\label{F:FGAfoo comparison}
\end{subfigure}
\caption{Target property comparisons for FACE A and surrogates.  Shown are surrogates developed in this work and past literature efforts. Ahmed et al.~\cite{Ahmed2015} did not consider MON in their surrogate formulations for FACE A}
\label{fig:facea properties}
\end{figure}

Table~\ref{T:FACE_A} compares the performances of the full-palette, eight-component and reduced-palette, four-component surrogates developed in the current study for FACE gasoline A along with the five-component surrogates developed by Sarathy et al.~\cite{Sarathy2014} and Ahmed et al.~\cite{Ahmed2015}.  For clarity, Ahmed et al. presents three FACE A surrogates; however, one of them is from Sarathy et al.~\cite{Sarathy2014} for comparison.   
Figure~\ref{F:FGApiona comparison} shows that the full-palette surrogate well-matches the n-paraffins and isoparaffins hydrocarbon classes in FACE gasoline A. The aromatic content of this surrogate surmounts to the the combined content of olefins, naphthenes, and aromatics.
The proposed surrogates from the literature match well, but do not account for the aromatics, olefins, or nephthenes.
The reduced-palette surrogate comprises only isoparaffins and aromatics, where the \textit{n}-paraffin content has been replaced by isoparaffins. 
Figure~\ref{F:FGAcarbon comparison} shows the \ce{C-C} bond type proportions; both surrogates developed in the current study closely match FACE A in all groups, despite the differences in overall hydrocarbon class makeup.
Our surrogates better match the distillation characteristics of FACE A, with a higher T$_{b}$ above \SI{60}{\percent} distillate than the literature surrogates as Figure~\ref{F:FGAdist comparison} shows.
We attribute this to the higher boiling components, which other proposed surrogates do not contain.
We find matching higher T$_{b}$ comes at the expense of more error with formulated PIONA, since our aromatic composition in the surrogate are larger than the combined aromatic, olefin, and naphthenes of the target fuel.  Future efforts will investigate if this is due to ADC modeling artifacts, discrepancies in ADC to ASTM-86 methods, and/or if it is preferential to match PIONA versus higher T$_{b}$ targets in the surrogate formulations. 
Lastly, Figure~\ref{F:FGAfoo comparison} shows the error between the remaining target properties---recall that the development of surrogates proposed by Ahmed et al.~\cite{Ahmed2015} for FACE A did not consider MON.
The current surrogates match RON better than the literature surrogates, but at the expense of matching H/C and density.
The full-palette surrogate slightly outperforms the reduced-palette surrogate in these four final metrics.

\subsubsection{FACE C}

\begin{table}[htbp]
\centering
\footnotesize
\begin{tabular}{@{}lccccc@{}}
\toprule
%& & & & \multicolumn{4}{c}{\footnotesize{FACE C Surrogates}}\\
%\cmidrule{3-7}\cmidrule{9-12}
%
Parameter 	& FACE C\cite{Cannella:2014aa} & Full & Reduced & ~~Sarathy\cite{Sarathy2014} &~~Ahmed\cite{Ahmed2015} \\
\midrule
RON & 84.7 & 84.1 &	84.4 & 84.0 & 85.3\\
MON	& 83.5 & 83.2 &	82.9 & 84.0 & \\
Density	[\si{\kilo\gram\per\meter^3}] & 690 & 703 & 698 & 686 & 696 \\
H/C	& 2.27 & 2.16 &	2.20 & 2.25 & 2.23 \\
\midrule
\% Volume distilled	& \multicolumn{5}{c}{Temperature [\si{\kelvin}]} \\
\midrule
10	& 331 & 336 & 334 & 325 & 329 \\
20  & 341 & 341 & 340 & 341 & 344\\
30	& 350 & 348 & 348 & 354 & 357\\
40  & 359 & 355 & 358 & 363 & 365\\
50  & 366 & 363 & 367 & 367 & 368\\
60  & 372 & 370 & 372 & 369 & 370\\
70  & 376 & 374 & 375 & 370 & 372\\
80  & 382 & 378 & 378 & 370 & 374\\
\midrule
Species & \multicolumn{5}{c}{Molar \%} \\
\midrule
\textit{n}-butane   &   &     &       & 17.0 & 18.4\\
\textit{n}-pentane	&   & 5.6 & 0     & & \\
\textit{n}-heptane	&   & 5.2 & 13.0   & 11.0 & 12.5\\
2-methylbutane	    &   & 18.6 & 28.2 & 8.0 & 7.0\\
2-methylpentane		&   & 13.3 & 5.5  & & \\
2-methylhexane	    &   & 6.5  & 0    & 5.0 & 4.7 \\
2,2,4-trimethylpentane  & & 40.0 & 45.9 & 56.0 & 54.6 \\
toluene	            &   & 4.3 & 0     & 3.0 & 4.8 \\
\textit{o}-xylene   &   & 5.0 & 7.4   & & \\
1,2,4-trimethylbenzene& & 1.5 & 0     & & \\
\bottomrule
\end{tabular}
\caption{
The full- and reduced-palette FACE gasoline C surrogates compared with literature surrogates and the real FACE C properties. 
A blank entry indicates the species\slash parameter was not considered. A zero (0) indicates the species was in the palette, but not chosen by the optimizer.
}
\label{T:FACE_C}
\end{table}

\begin{figure}[htbp]
\centering
\begin{subfigure}[b]{.5\linewidth}
\includegraphics[width=85mm]{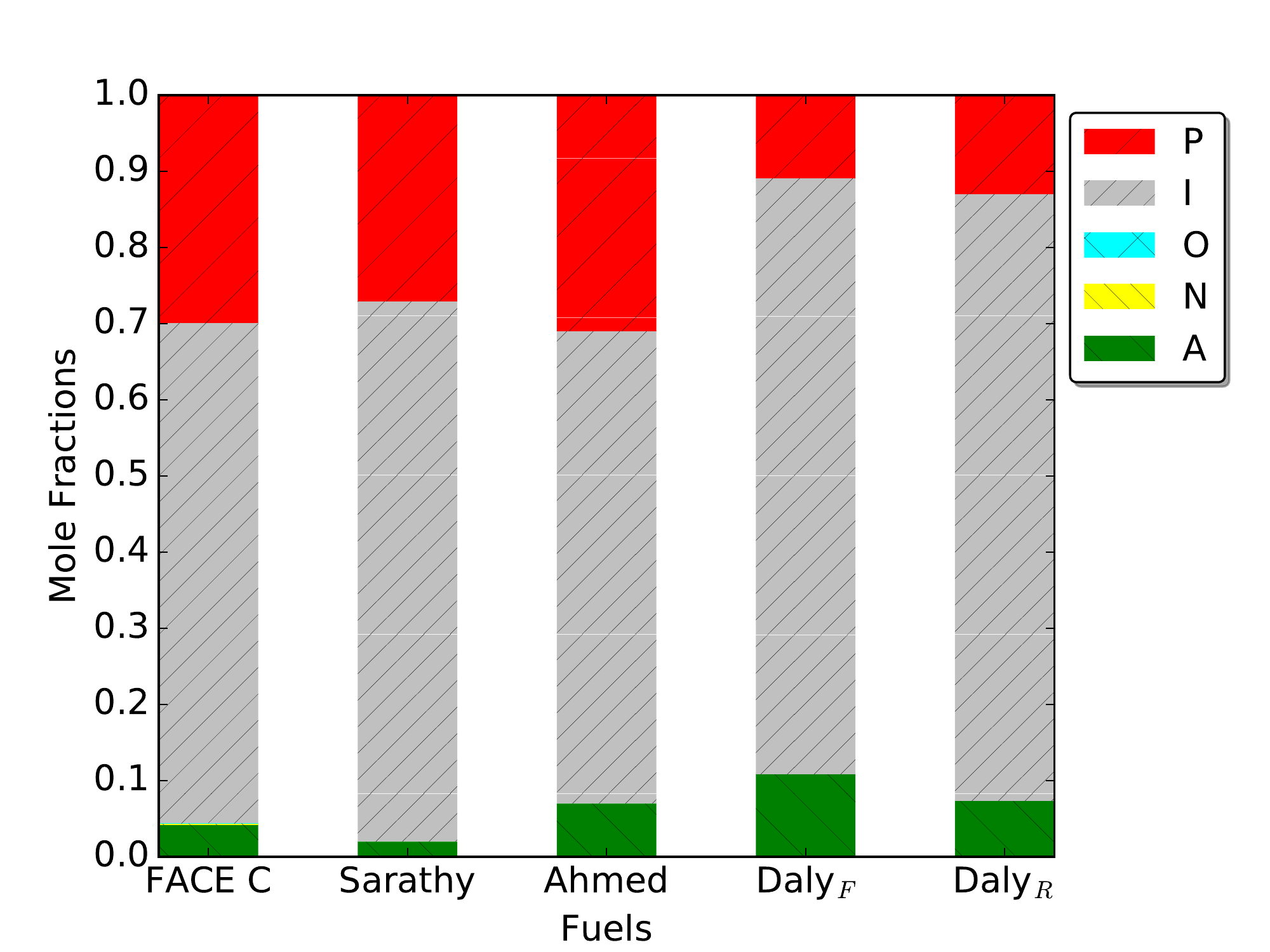}
\caption{Hydrocarbon class proportions}\label{F:FGCpionaa}
\end{subfigure}
~
\begin{subfigure}[b]{.45\linewidth}
\includegraphics[width=85mm]{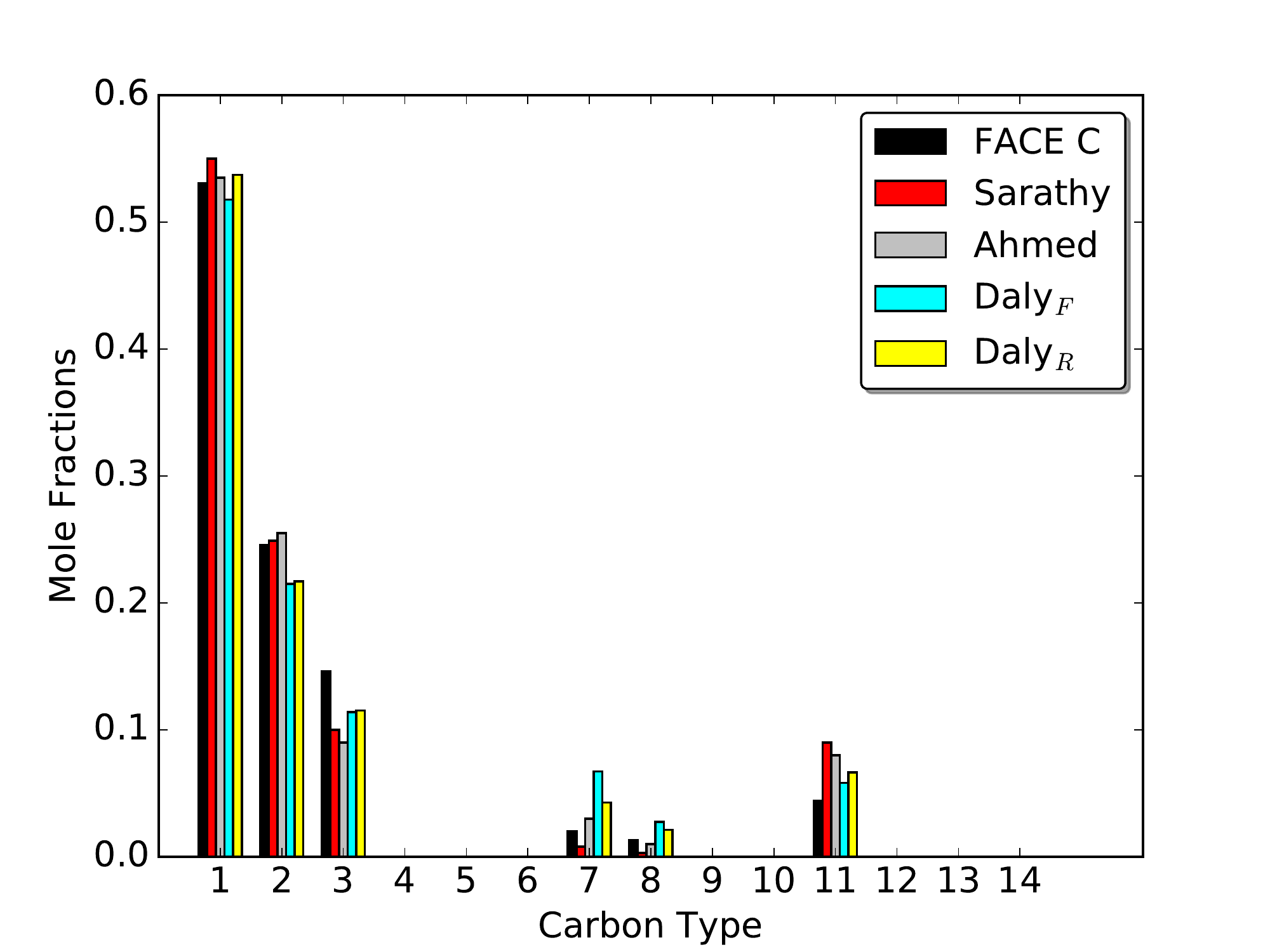}
\caption{C--C bond type proportions}\label{F:FGCcarbona}
\end{subfigure}
\\
\begin{subfigure}[b]{.5\linewidth}
\includegraphics[width=85mm]{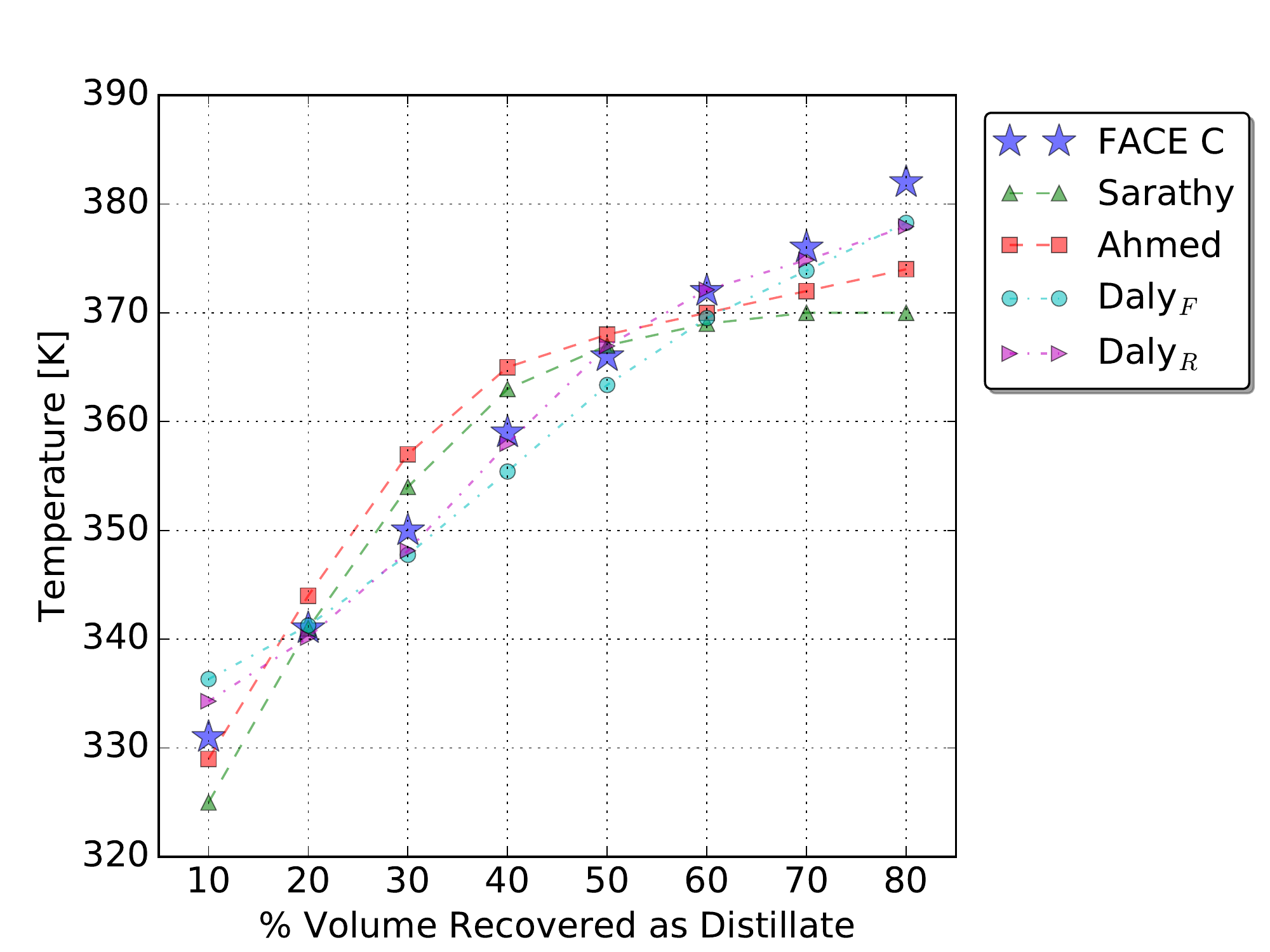}
\caption{Distillation characteristics}\label{F:FGCdista}
\end{subfigure}
~
\begin{subfigure}[b]{.45\linewidth}
\includegraphics[width=85mm]{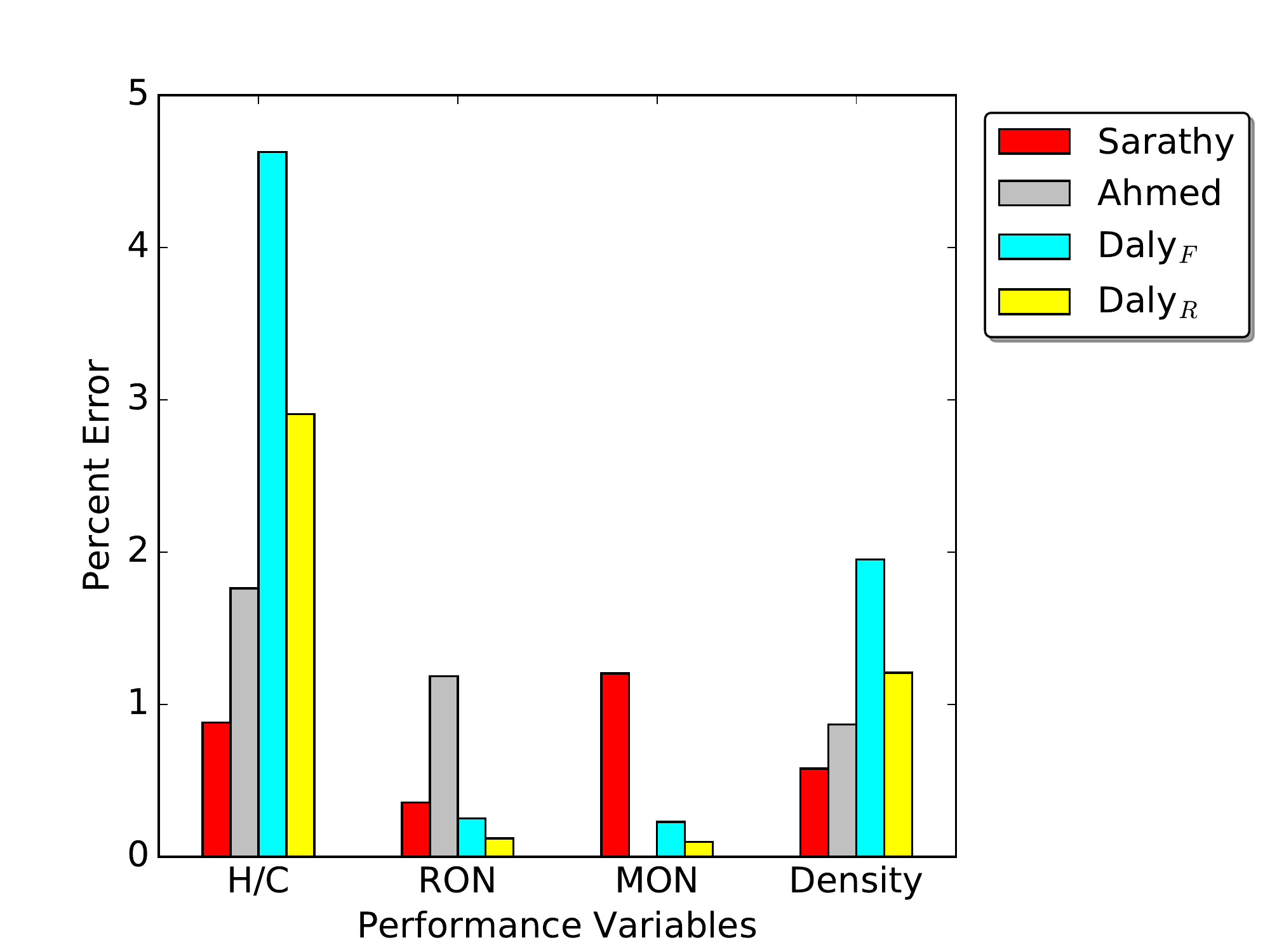}
\caption{Error in H/C, RON, MON, and density}\label{F:FGCfooa}
\end{subfigure}
\caption{Target property comparisons for FACE C and surrogates.  Shown are surrogates developed in this work and past literature efforts.  Ahmed et al.~\cite{Ahmed2015} did not consider MON in their surrogate formulation for FACE C}
\label{fig:facec properties}
\end{figure}

Table~\ref{T:FACE_C} compares the FACE gasoline C surrogates with the six-component surrogates of Sarathy et al.~\cite{Sarathy2014} and Ahmed et al.~\cite{Ahmed2015}.  Again, Ahmed et al.\ presents two FACE C surrogates but one of them is from Sarathy et al.~\cite{Sarathy2014} for comparison. 
Our full palette consists of a nine-component surrogate, as well as a reduced four-component version.
It can be seen in Figure~\ref{F:FGCpionaa}, that the full- and reduced-palette surrogate has less \textit{n}-paraffins and more aromatic content than those by Ahmed et al. 
The reduced-palette surrogate, in comparison with the reduced surrogate for FACE A, retains \textit{n}-paraffins content.
The \ce{C-C} bond type proportions are again seen in Figure~\ref{F:FGCcarbona}, showing a trade-off with all surrogates for the \ce{C-C} groupings; no surrogate matches all groups perfectly.
The \ce{C-C} proportions remain similar, with a small trade-off between bond groups such as 1 and 7, in congruence with PIONA.
Figure~\ref{F:FGCdista} presents the distillation curves.
The current surrogates show higher T$_{b}$, relative to literature surrogates, above \SI{60}{\percent} distillate to better match FACE C.
We attribute this to higher compositions of the higher boiling components over the other surrogates.
Again, we find that matching higher T$_{b}$ comes at the expense of an arguably less-optimal solution for PIONA.  As discussed with FACE A surrogates, future efforts aim to investigate these findings in more detail.   
Interestingly, the reduced surrogate matches the target curve best at all distillate percentages out of all surrogates (except at \SI{10}{\percent}).
Lastly, Figure~\ref{F:FGCfooa} shows the error between the remaining target properties---recall surrogates proposed in Ahmed et al.~\cite{Ahmed2015} for C did not consider MON.
We can see that the RON are better-matched in both new surrogates, but at the expense of H/C and density.

\subsubsection{FACE F}

\begin{table}[htbp]
\footnotesize
\centering
\begin{tabular}{@{}l c c c c c@{}}
\toprule
Parameter 	& FACE F\cite{Cannella:2014aa} & ~~Full~~ & ~~Reduced~ & KAUST\cite{Sarathy2016} & LLNL\cite{Sarathy2016} \\
\midrule
RON 	        &94.4 & 93.6 &94.3	 & 93.6 & 93.8 \\
MON		        &88.8 & 88.2 &88.7	 & 88.9 & 89.5 \\
Density	[\si{\kilo\gram\per\m^3}] & 707  & 734 & 731.8 & 707 & 712 \\
H/C				&2.13 & 2.03 &	2.03 & 2.12 & 2.06 \\
\midrule
\% Volume distilled	& \multicolumn{5}{c}{Temperature [\si{\kelvin}]}  \\
\midrule
10			& 346 & 345 &   346	& 350 &353\\
20		    & 351 & 350 &   351	& 356 &356\\
30	        & 357 & 356 &   357 & 361 &358\\
40	        & 363 & 363 &	363	& 366 &361\\
50	        & 370 & 367 &	369	& 370 &364\\
60	        & 376 & 375 &	374	& 374 &366\\
70	        & 382 & 380 &	379	& 378 &369\\
80	        & 387 & 387 &	386	& 383 &371\\
\midrule
Species &  \multicolumn{5}{c}{Molar \%} \\
\midrule
\textit{n}-butane       & & & & 6.9 & 0 \\
\textit{n}-pentane 	    & & 0 & 0 & & \\
\textit{n}-heptane	    & & 0 & 0 & 0 & 7.0 \\
2-methylbutane	        & & 9.4 & 9.4 & 9.8 & 0 \\
2-methylpentane		    & & 0.1 & 0 & & \\
2-methylhexane	        & & 10.4 & 7.5 & 7.0 & 0 \\
2,2,4-trimethylpentane	& & 38.9 & 41.5 & 43.7 & 53.0 \\
1-pentene	            & & 0 & 0 & &  \\
1-hexene	            & & 8.3 & 11.6 & 8.4 & 14.0 \\
cyclopentane	        & & 16.5 & 14.3 & 15.8 & 14.0 \\
cyclohexane             & & 0 & 0 & & \\
toluene	                & & 0 & 0 & 0 & 12.0 \\
\textit{o}-xylene		& & 16.4 & 15.7 & & \\
1,2,4-trimethylbenzene	& & 0 & 0 & 8.4 & 0 \\
\bottomrule
\end{tabular}
\caption{
The full- and reduced FACE F surrogates compared with literature surrogates and the real FACE F properties. A blank entry indicates the species\slash parameter was not considered. A zero (0) indicates the species was in the palette, but not chosen by the optimizer.  
}
\label{T:FACE_F}
\end{table}

\begin{figure}[htbp]
\centering
\begin{subfigure}[b]{.5\linewidth}
\includegraphics[width=85mm]{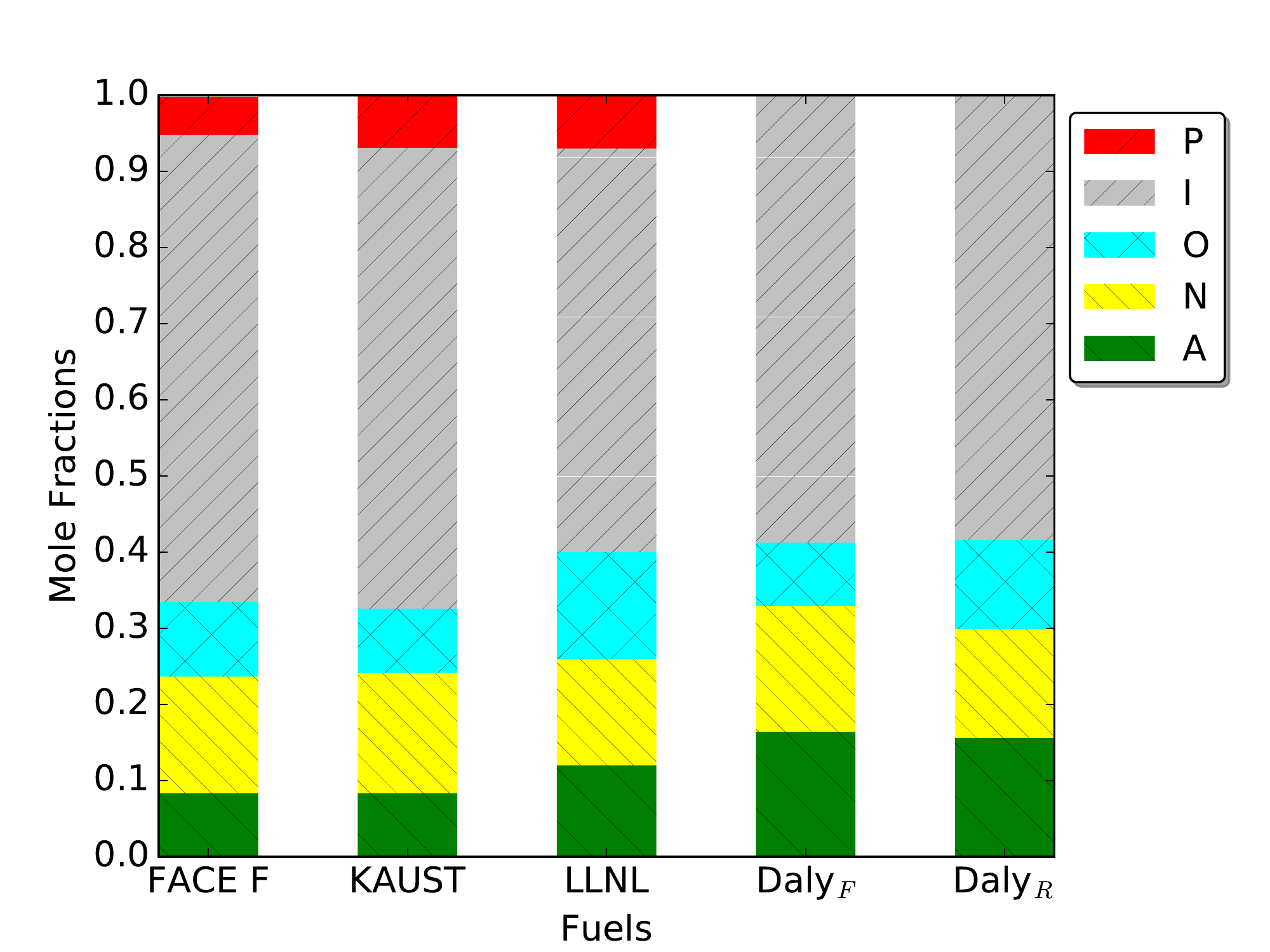}
\caption{Hydrocarbon class proportions}\label{F:FGFpionaa}
\end{subfigure}
~
\begin{subfigure}[b]{.45\linewidth}
\includegraphics[width=85mm]{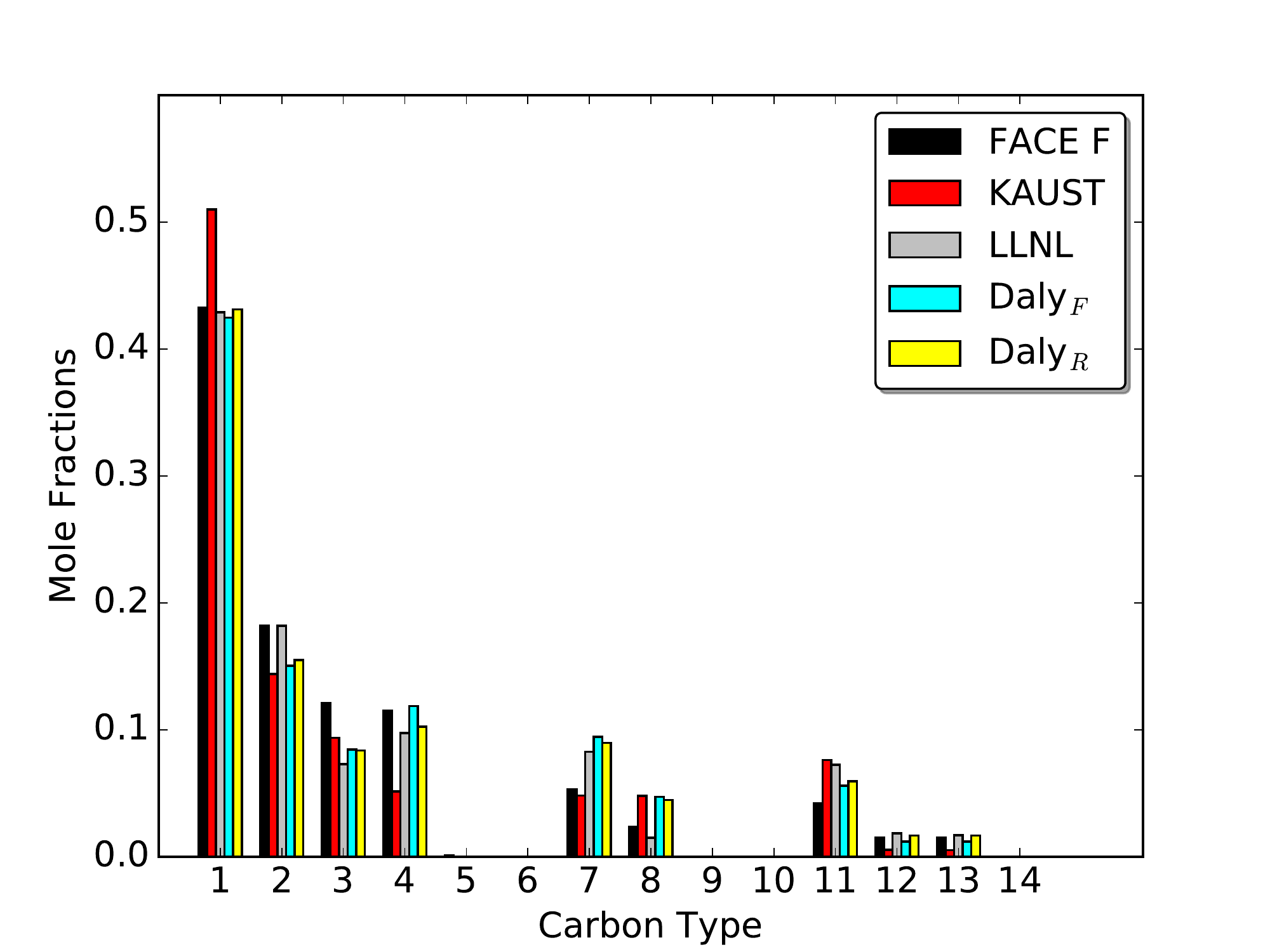}
\caption{C--C bond type proportions}\label{F:FGFcarbona}
\end{subfigure}
\\
\begin{subfigure}[b]{.5\linewidth}
\includegraphics[width=85mm]{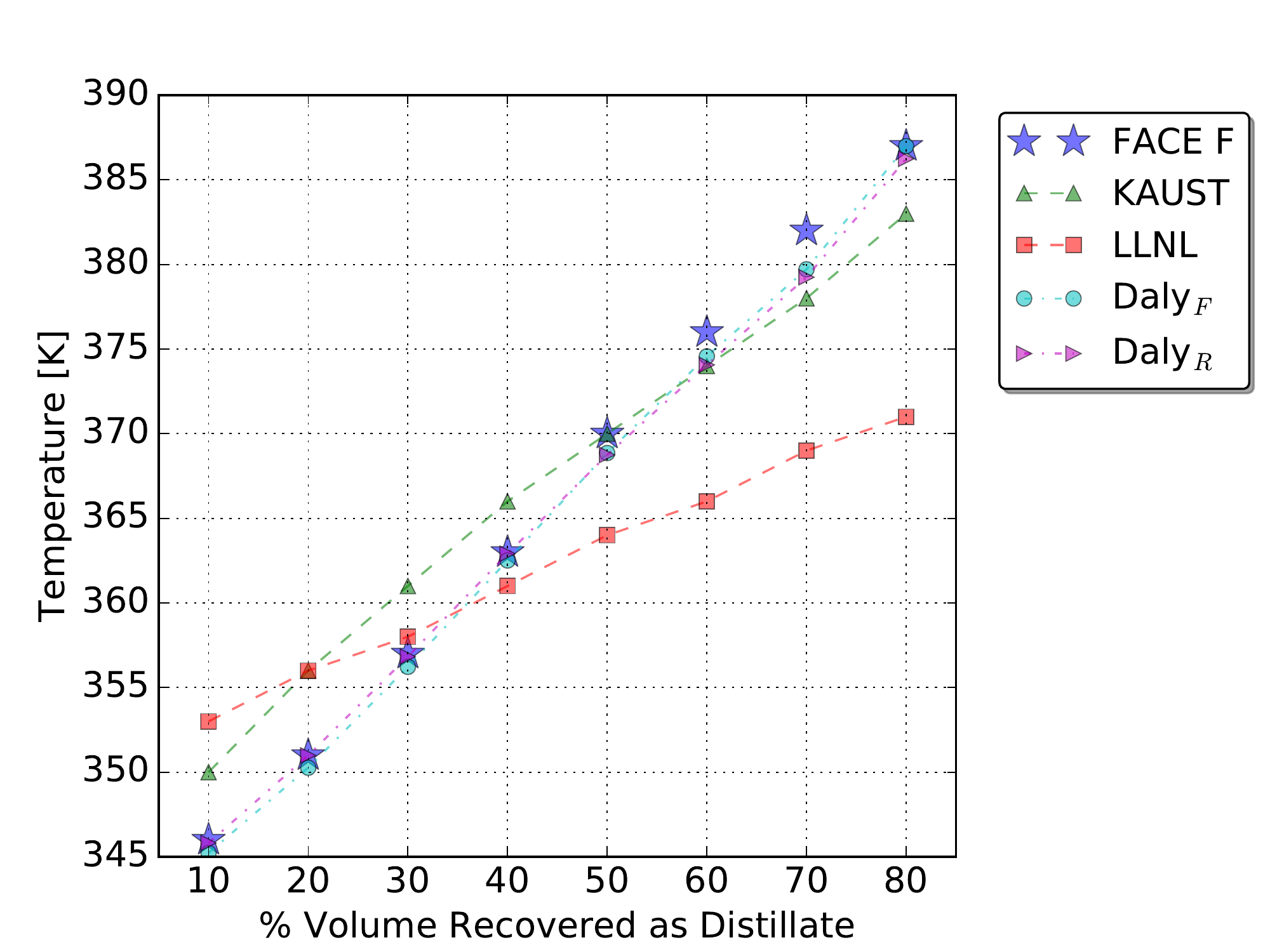}
\caption{Distillation curve}\label{F:FGFdista}
\end{subfigure}
~
\begin{subfigure}[b]{.45\linewidth}
\includegraphics[width=85mm]{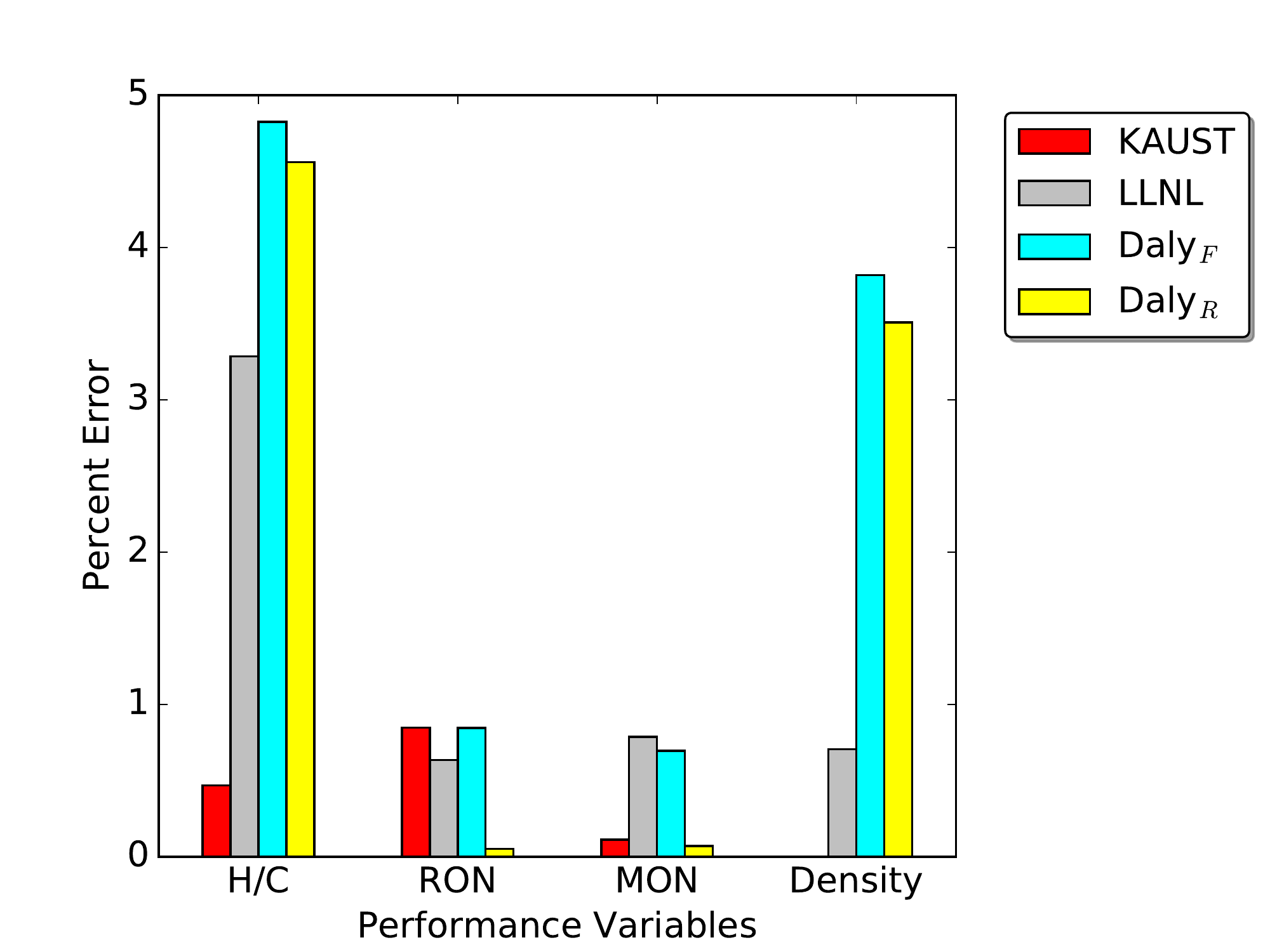}
\caption{Error in H/C, RON, MON, and density}\label{F:FGFfooa}
\end{subfigure}
\caption{Target property comparisons for FACE F and surrogates.  Shown are surrogates developed in this work and past literature efforts.}
\label{fig:face F properties}
\end{figure}

FACE gasoline F surrogates are compared in Table~\ref{T:FACE_F} with those presented in Sarathy et al.~\cite{Sarathy2016}, which utilized eight and seven components.
We present a seven-component surrogate with alternate isoparaffins and additional aromatics to those in Sarathy et al.~\cite{Sarathy2016}, as well as a reduced six-component version.
It can be seen in Figure~\ref{F:FGFpionaa}, that the full surrogate has no \textit{n}-paraffins, more aromatics, and similar olefinic and naphthenic content than FACE F.
The PIONA of the full surrogate are similar to that of Sarathy et al.~\cite{Sarathy2016}, with the exception of \textit{n}-paraffins.
The surrogates lacking \textit{n}-paraffins are likely attributable to the IR-RON and MON correlation attempting to match octane sensitivity.
As found with the FACE C surrogates, Figure~\ref{F:FGFcarbona} shows a trade-off with all surrogates for the \ce{C-C} groupings, and no surrogate matches all groups completely.
Once again, the \ce{C-C} proportions remain similar in accordance with PIONA for both surrogates. 
Figure~\ref{F:FGFdista} shows the Daly surrogates have higher T$_{b}$ above \SI{60}{\percent} distillate to better match FACE F.
We attribute this to the increased amount of higher boiling components in comparison to the other surrogates.
Both full and reduced surrogates have nearly the same distillate curve with little error in comparison to the target fuel.
Lastly, Figure~\ref{F:FGFfooa} shows the error between the remaining target properties---Sarathy's surrogates for~\cite{Sarathy2016}FACE F did consider MON.
We can see that the reduced surrogate outperforms the full surrogate across the board, and better-matches RON and MON in comparison to all other surrogates.
Both our full and reduced surrogates do not match H/C or density as well as those from Sarathy et al.~\cite{Sarathy2016}.

The octane ratings of the LLNL FACE F surrogate, as predicted by our FTIR-octane model, surmount to a RON of 91.4 and MON of 86.0.  These are not in agreement with the reported values of 93.8 and 89.5 for RON and MON, respectively.  Sarathy et al. did not perform auto-ignition or RON and MON tests for the LLNL surrogates~\cite{Sarathy2016} (but did for KAUST surrogates), so at this time the octane correlation discrepancies are left unresolved and a study for future efforts.  As previously discussed, we cannot evaluate the KAUST surrogates because our FTIR-octane model does not incorporate n-butane.

\subsubsection{FACE G}

\begin{table}[htbp]
\footnotesize
\centering
\begin{tabular}{@{}l c c c c c@{}}
\toprule
Parameter 	& ~~FACE G\cite{Cannella:2014aa} & ~~Full~~ & ~Reduced~ & ~~KAUST\cite{Sarathy2016} & ~~LLNL\cite{Sarathy2016} \\
\midrule
RON 	    & 96.8 & 95.9 &	96.3 & 95.2 & 96.4 \\
MON		    & 85.8 & 86.7 &	86.8 & 87.9 & 85.5 \\
Density [\si{\kilo\gram\per\m^3}] & 760  & 803 & 802 & 742 & 751  \\
H/C ratio   & 1.83 & 1.57 &	1.54 &	1.85  & 1.87 \\
\midrule
\% Volume distilled	& \multicolumn{5}{c}{Temperature [\si{\kelvin}]}  \\
\midrule
10			& 350& 348& 352&345 &361 \\
20		    & 363& 358& 363&351 &366 \\
30	        & 378& 372& 374&359 &371 \\
40	        & 394& 389& 387&367 &378 \\
50	        & 411& 406& 404&376 &385 \\
60	        & 426& 418& 419&386 &393 \\
70	        & 439& 423& 424&397 &405 \\
80	        & 447& 426& 426&417 &426 \\
\midrule
Species & \multicolumn{5}{c}{Molar \%} \\
\midrule
\textit{n}-butane   & &  & & 7.6 & 0 \\
\textit{n}-pentane 	& & 0 & 0 & & \\
\textit{n}-heptane  & & 3.9 & 0 & 0 & 8.0 \\
2-methylbutane	    & & 0.0 & 0 & 9.5 & 0 \\
2-methylpentane		& & 0.5 & 0 & & \\
2-methylhexane	    & & 0 & 0 & 9.8 & 0 \\
2,2,4-trimethylpentane	& & 1.8 & 0 & 18.0 & 38.0\\
1-pentene	        & & 12.1 & 9.5 & & \\
1-hexene	        & & 20.1 & 35.7 & 8.1 & 9.0 \\
cyclopentane	    & & 9.3 & 0  & 15.3 & 14.0 \\
cyclohexane         & & 0 & 0 & & \\
toluene	            & & 0 & 0 & 10.6 & 0\\
\textit{o}-xylene   & & 38.9 & 40.0 & & \\
1,2,4-trimethylbenzene	& & 14.4 & 14.8 & 21.1 & 31.0\\
\bottomrule
\end{tabular}
\caption{
The full- and reduced FACE G surrogates compared with literature surrogates and the real FACE G properties. A blank entry indicates the species\slash parameter was not considered. A zero (0) indicates the species was in the palette, but not chosen by the optimizer.
}
\label{T:FACE_G}
\end{table}

\begin{figure}[htbp]
\centering
\begin{subfigure}[b]{.5\linewidth}
\includegraphics[width=85mm]{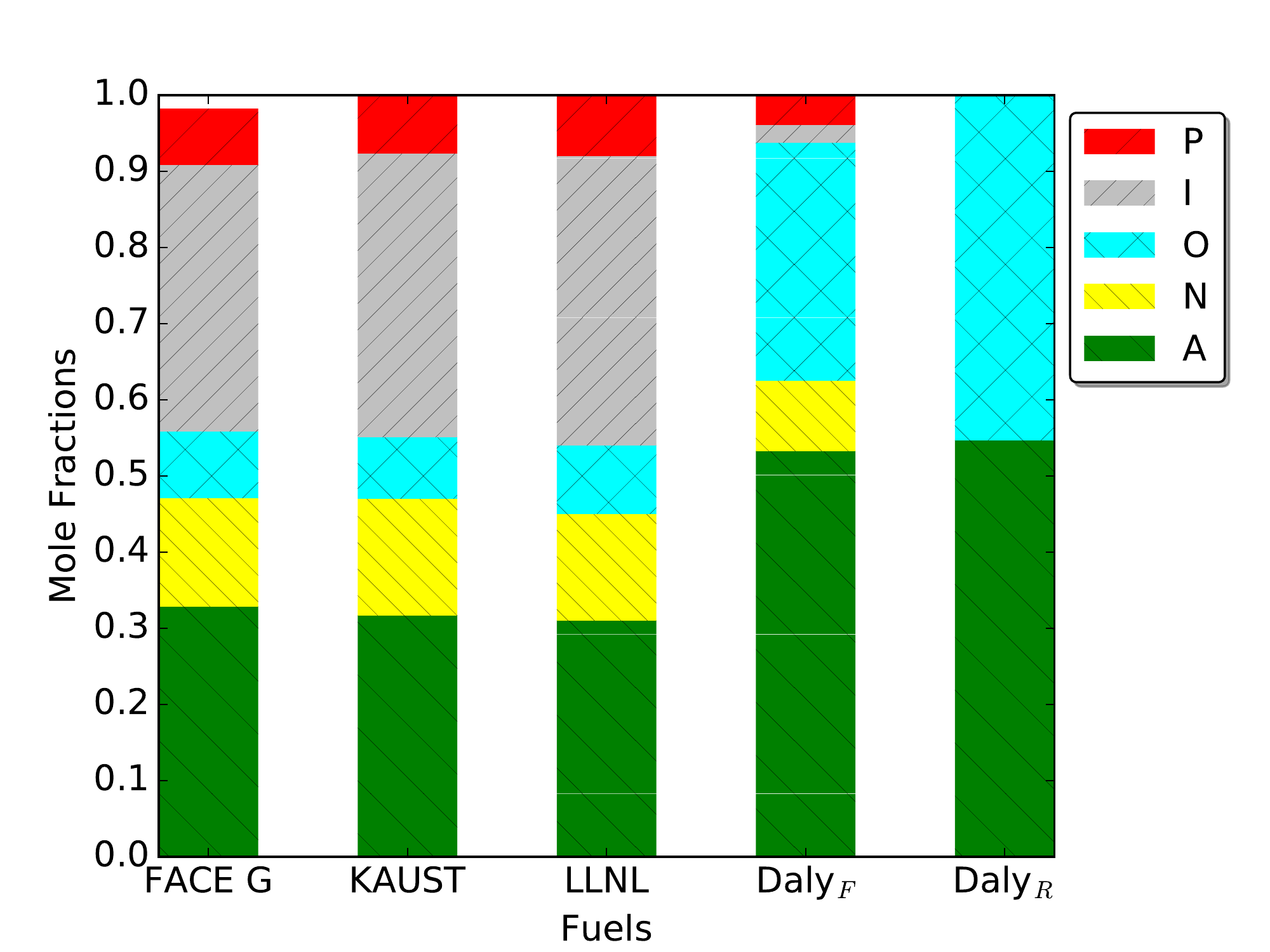}
\caption{Hydrocarbon class proportions}\label{F:FGGpionaa}
\end{subfigure}
~
\begin{subfigure}[b]{.45\linewidth}
\includegraphics[width=85mm]{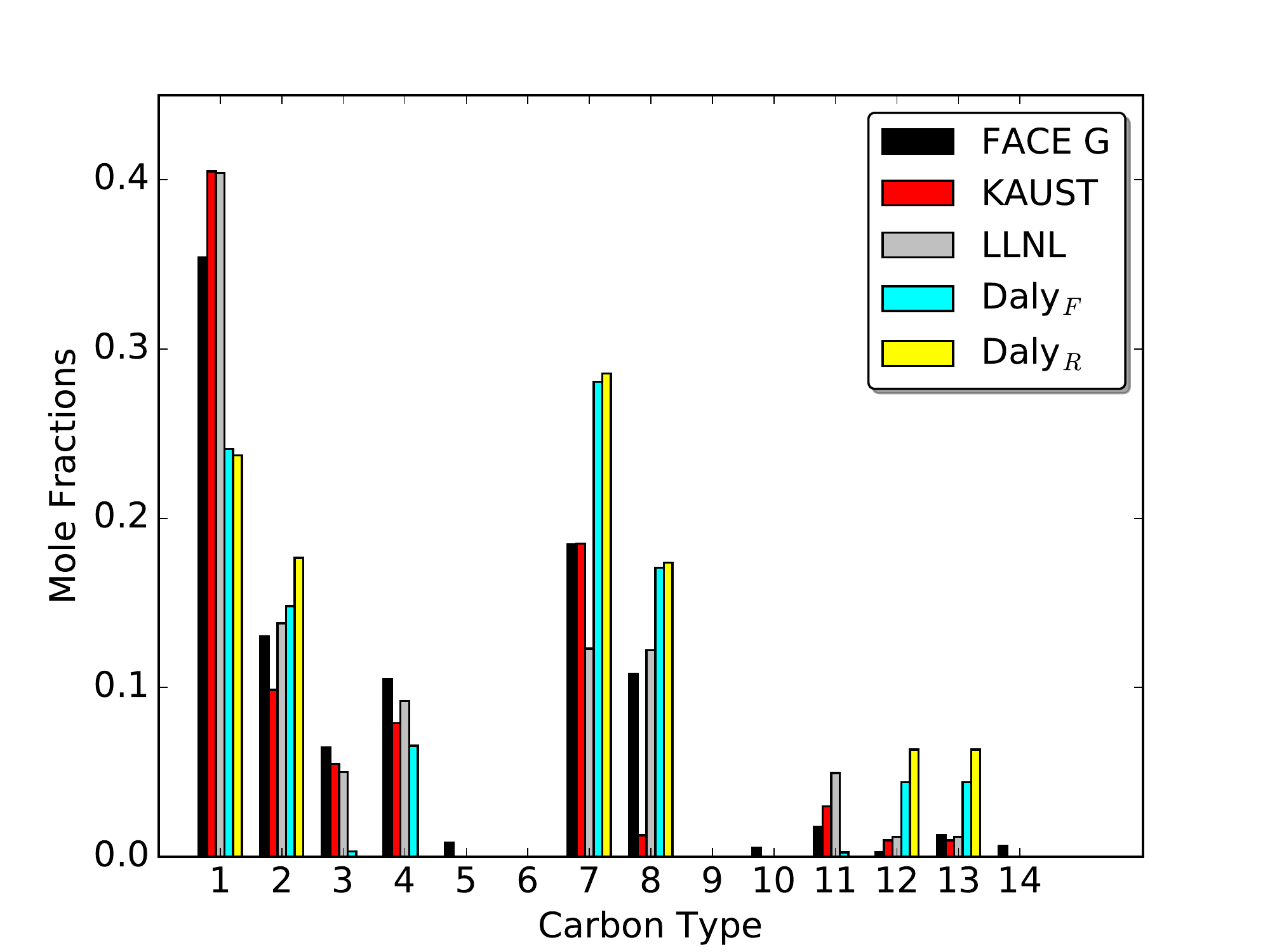}
\caption{C--C bond type proportions}\label{F:FGGcarbona}
\end{subfigure}
\\
\begin{subfigure}[b]{.5\linewidth}
\includegraphics[width=85mm]{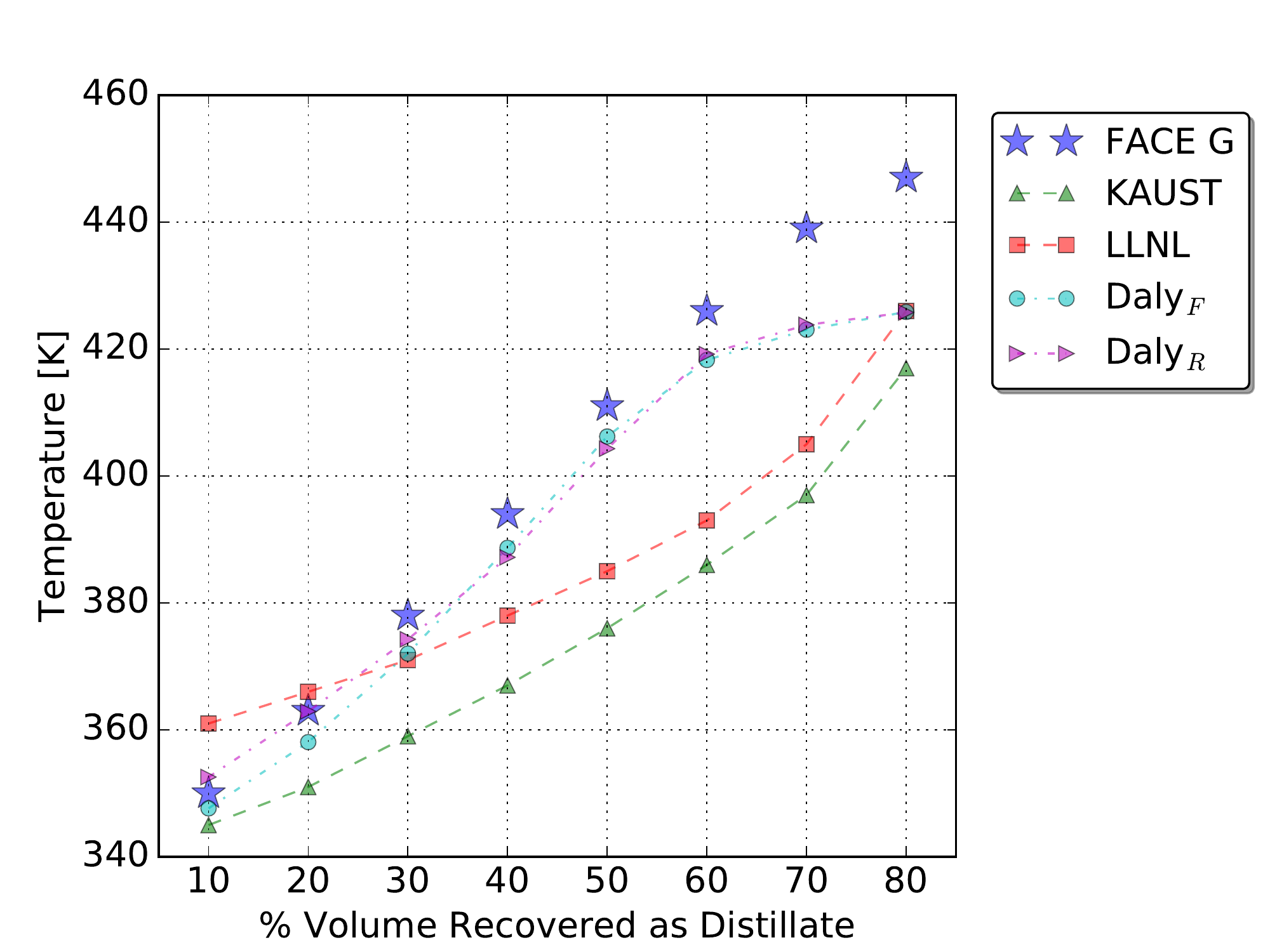}
\caption{Distillation Characteristics}\label{F:FGGdista}
\end{subfigure}
~
\begin{subfigure}[b]{.45\linewidth}
\includegraphics[width=85mm]{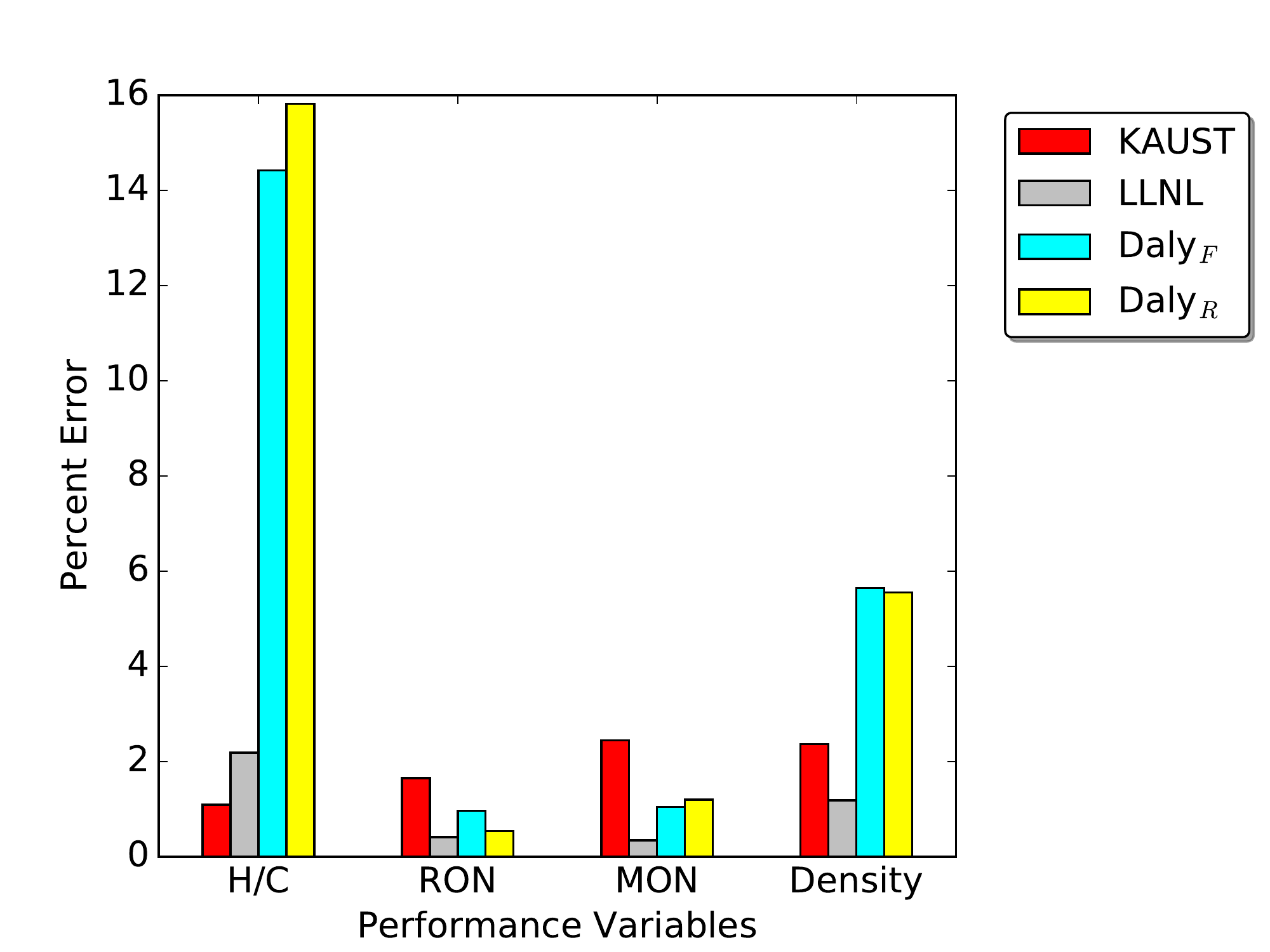}
\caption{Error in H/C, RON, MON, and density}\label{F:FGGfooa}
\end{subfigure}
\caption{Target property comparisons for FACE G and surrogates.  Shown are surrogates developed in this work and past literature efforts.}
\label{fig:face G properties}
\end{figure}

In Table~\ref{T:FACE_G}, FACE gasoline G surrogates are compared with those presented in Sarathy et al.~\cite{Sarathy2016}.
This work presents a full palette eight-component surrogate, as well as a reduced four-component version.
It can be seen in Figure~\ref{F:FGGpionaa}, that the full surrogate has more aromatics and olefins than FACE G, with less naphthenes, isoparaffins and \textit{n}-paraffins.
PIONA of the full surrogate are dissimilar to those proposed by Sarathy et al.~\cite{Sarathy2016}, which match the PIONA of FACE G very well.
The reduced surrogate is composed only of olefins, naphthenes, and aromatics, with the olefinic content mostly replacing the \textit{n}-paraffins and isoparaffins.
Figure~\ref{F:FGGcarbona} shows the \ce{C-C} groupings for the current surrogates being relatively large for carbon types 7 and 8, due to the high olefins and aromatics, with a low amount of carbon type 1, from lack of \textit{n}-paraffins and isoparaffins.
Figure~\ref{F:FGGdista} shows the new surrogates match FACE G T$_{b}$ from 10 to \SI{70}{\percent} evaporated, better than the other surrogates.
However, all surrogates fail to capture the high T$_{b}$ above \SI{70}{\percent}, as a result of all surrogates lacking enough proportions of higher boiling components.
Logically, it seems advisable to reformulate these surrogates for FACE G with additional constraints to use the higher boiling 1,2,4-trimethylbenzene, as opposed to \textit{o}-xylene.
However, it is seen that the other surrogates proposed in literature, which do include 21-31\% of this molecule, do not match this distillation region either.
We can see, in Figure~\ref{F:FGGfooa}, that the current surrogates have high H/C and density errors, but low errors for RON and MON.  

FACE G was particularly challenging to formulate a surrogate for. 
The contributing factors are: 1) the optimization routine not being strictly constrained to follow hydrocarbon class proportions to match the target fuel, 2) the IR-octane models converging to high olefinic contents in order to formulate a high octane sensitivity fuel, and 3) the objective function highly weighting octane more-so than H/C or density.
In short, the optimization routine blended a high sensitivity fuel (by way of high olefinic content), at the expense of properly matching H/C, density, and hydrocarbon class proportions.  To support this, we evaluated the LLNL FACE G surrogate with our FTIR-octane model to check whether we would predict a high octane sensitivity fuel.  We calculated the RON to be 90.2 and MON of 89.9, in contradistinction to the reported values of 96.4 and 85.5.
We conclude that the formulation framework utilizing the IR-octane models, in their current state, and the components chosen for the palette are inadequate to match fuels having S > 10, H/C > 1.8, and density <\SI{760}{\kilo\gram\per\meter^3}---in other words, low olefinic content fuels with high octane sensitivity.
The flexibility in species bounds was found necessary to match octane of FACE G with the IR models. Future efforts will involve improving the IR model so tighter species constraints can be made to better-match PIONA and octane simultaneously.

\subsubsection{FACE I}

\begin{table}[htbp]
\footnotesize
\centering
\begin{tabular}{@{}l c c c c@{}}
\toprule
Parameter 	& ~~FACE I\cite{Cannella:2014aa} & ~~Full~~ & ~Reduced~ & ~~Javed\cite{Javed2017} \\
\midrule
RON 	    & 70.2 & 70.2 &	70.1 & 70.7 \\
MON		    & 70.1 & 69.5 &	69.5 & 68.4 \\
Density [\si{\kilo\gram\per\m^3}] & 697  & 716 & 710 & 706  \\
H/C ratio   & 1.92 & 2.12 &	2.15 &	2.22  \\
\midrule
\% Volume distilled	& \multicolumn{4}{c}{Temperature [\si{\kelvin}]}  \\
\midrule
10			& 343& 343& 341&350 \\
20		    & 354& 348& 347&354 \\
30	        & 359& 354& 353&358 \\
40	        & 362& 360& 360&361 \\
50	        & 364& 367& 364&363 \\
60	        & 366& 370& 367&365 \\
70	        & 368& 373& 368&366 \\
80	        & 371& 378& 369&367 \\
\midrule
Species & \multicolumn{4}{c}{Molar \%} \\
\midrule
\textit{n}-butane   & &  & &  \\
\textit{n}-pentane 	& & 13.0 & 11.0 & \\
\textit{n}-heptane  & & 15.0 & 11.6 & 12.0 \\
2-methylbutane	    & & 7.7 & 12.8 & 11.0 \\
2-methylpentane		& & 4.2 &  & \\
2-methylhexane	    & & 7.0 & 20.0 & 27.0\\
2,2,4-trimethylpentane	& & 28.5 & 28.5 & 34.0\\
1-pentene	        & & 0.6 &  & \\
1-hexene	        & & 3.0 &  & 6.0\\
cyclopentane	    & & 0.0 &  & 6.0\\
cyclohexane         & & 9.0 & 4.7 & \\
toluene	            & & 5.1 &  11.4& \\
\textit{o}-xylene   & & 3.3 & & \\
1,2,4-trimethylbenzene	& & 3.7 & & 4.0 \\
\bottomrule
\end{tabular}
\caption{
The full- and reduced FACE I surrogates compared with literature surrogates and the real FACE I properties. A blank entry indicates the species\slash parameter was not considered. A zero (0) indicates the species was in the palette, but not chosen by the optimizer.
}
\label{T:FACE_I}
\end{table}

\begin{figure}[htbp]
\centering
\begin{subfigure}[b]{.5\linewidth}
\includegraphics[width=85mm]{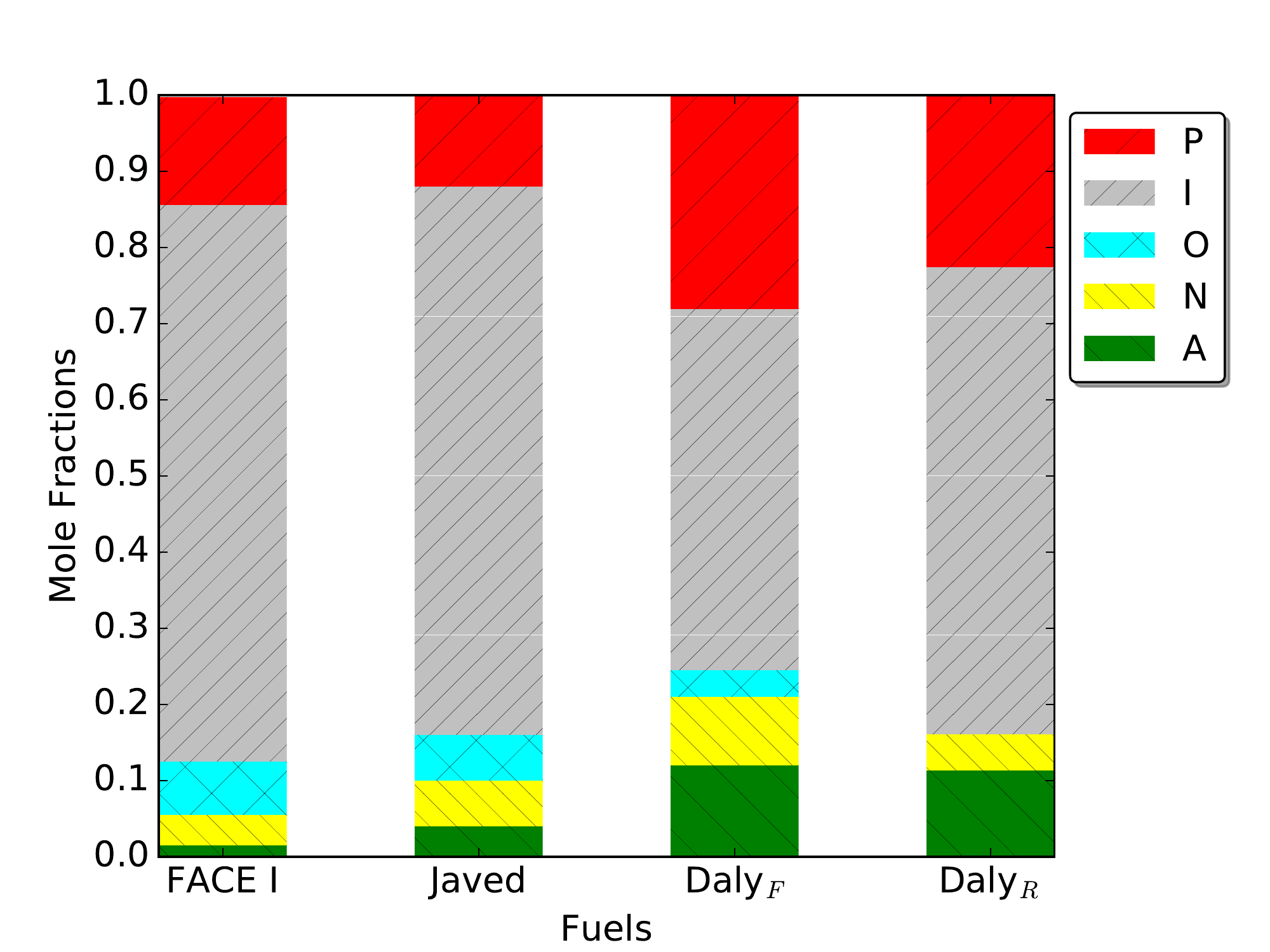}
\caption{Hydrocarbon class proportions}\label{F:FGIpionaa}
\end{subfigure}
~
\begin{subfigure}[b]{.45\linewidth}
\includegraphics[width=85mm]{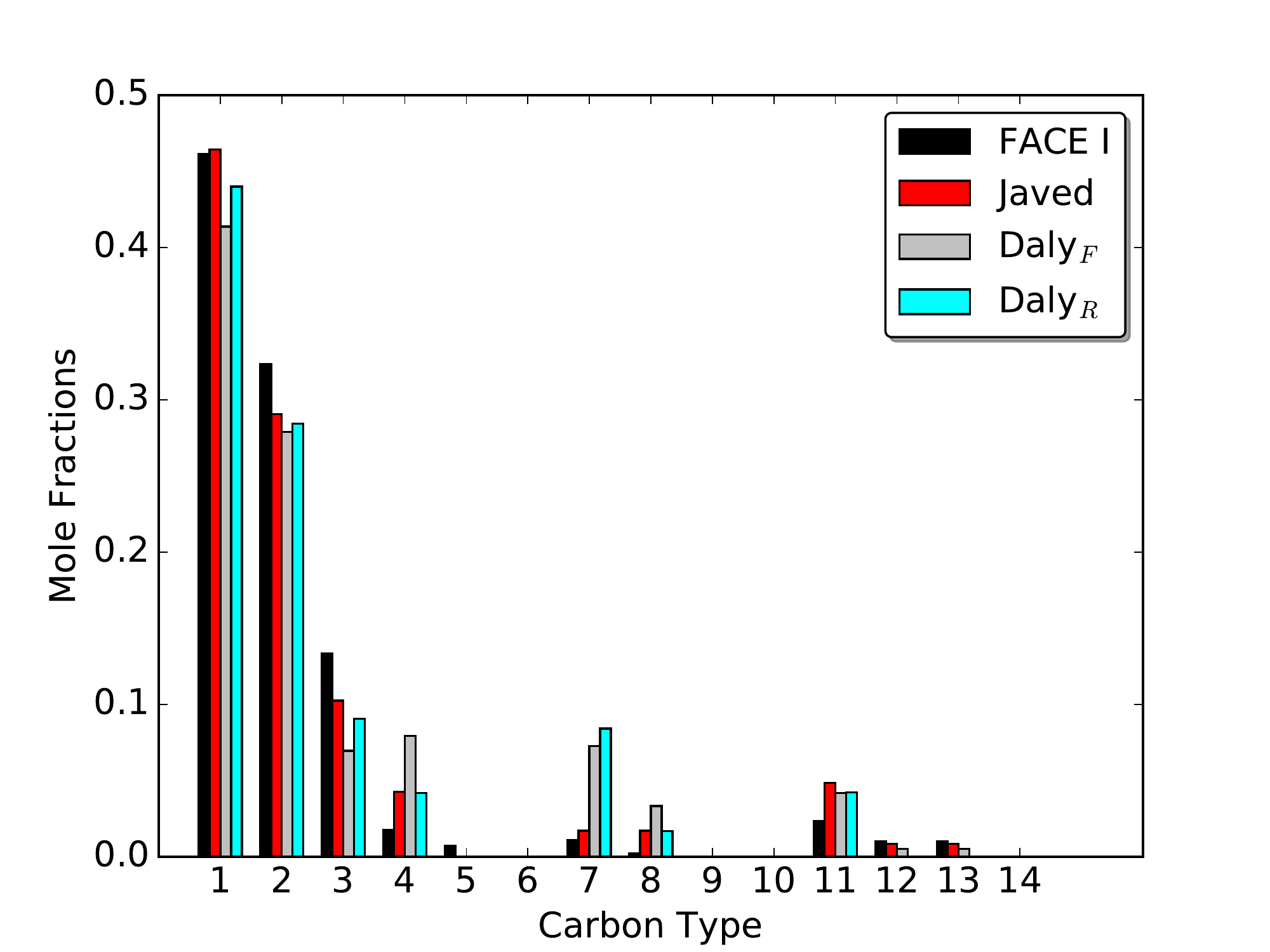}
\caption{C--C bond type proportions}\label{F:FGIcarbona}
\end{subfigure}
\\
\begin{subfigure}[b]{.5\linewidth}
\includegraphics[width=85mm]{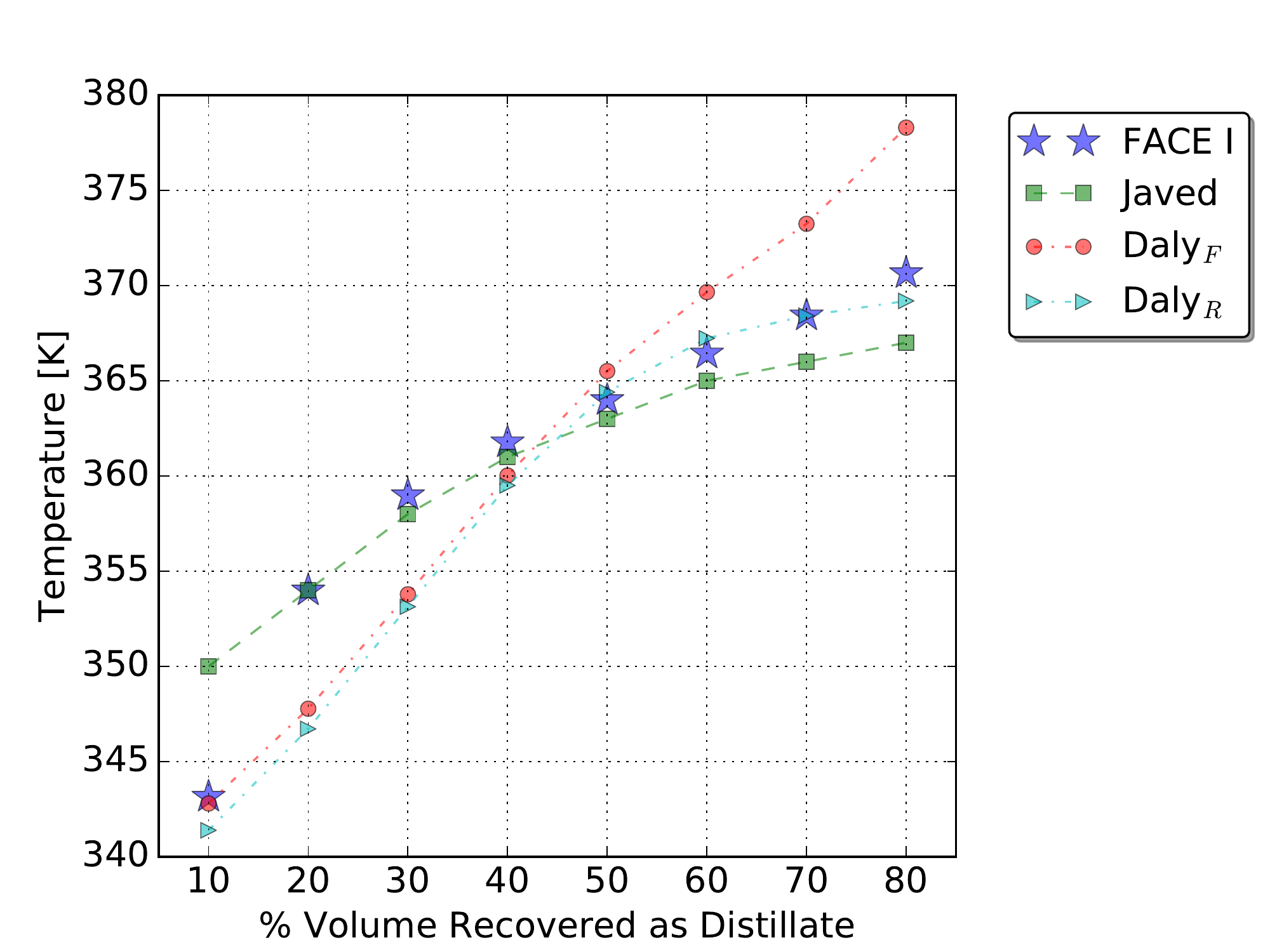}
\caption{Distillation Characteristics}\label{F:FGIdista}
\end{subfigure}
~
\begin{subfigure}[b]{.45\linewidth}
\includegraphics[width=85mm]{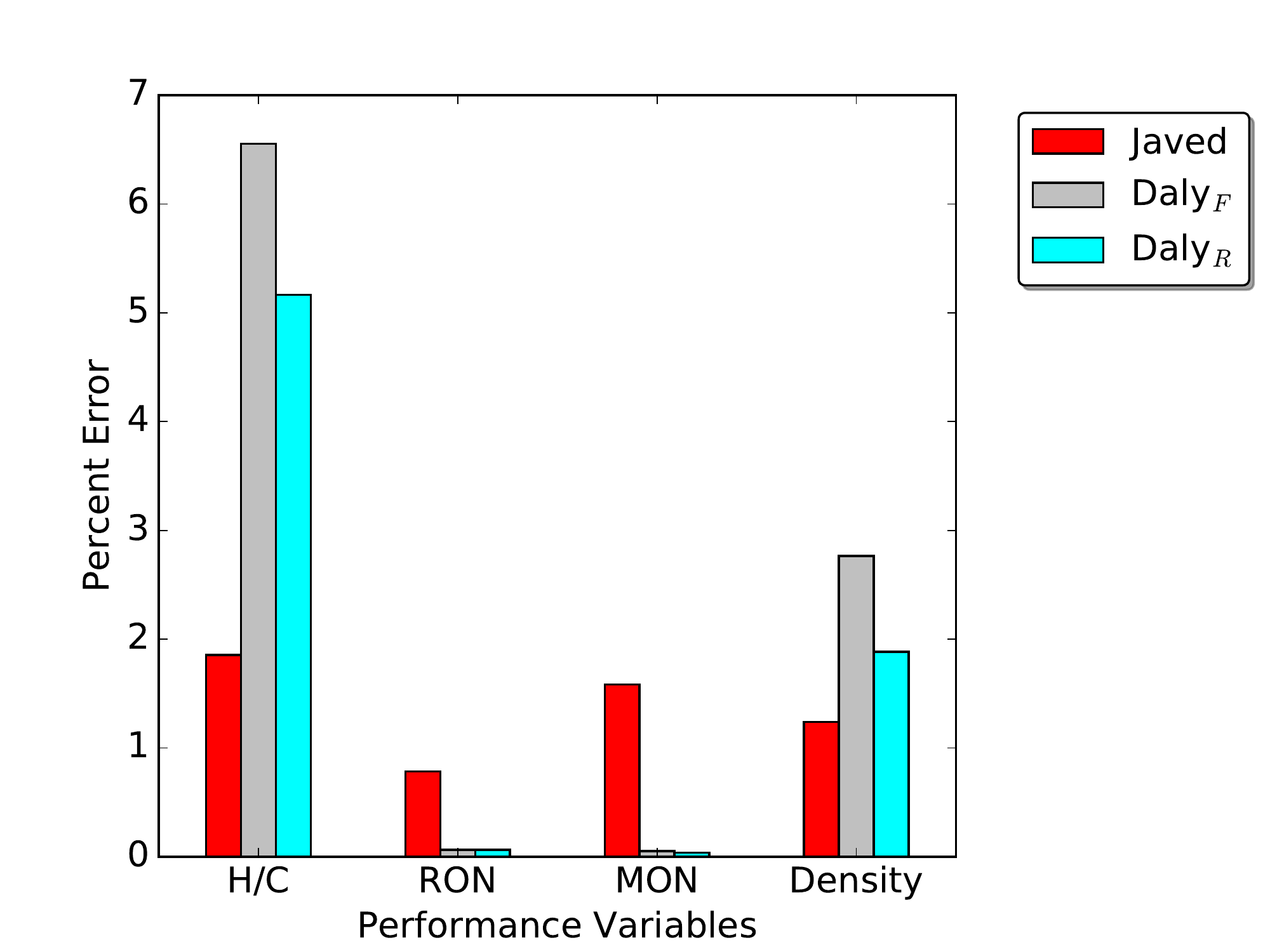}
\caption{Error in H/C, RON, MON, and density}\label{F:FGIfooa}
\end{subfigure}
\caption{Target property comparisons for FACE I and surrogates.  Shown are surrogates developed in this work and past literature efforts.}
\label{fig:face I properties}
\end{figure}

In Table~\ref{T:FACE_I}, FACE gasoline I surrogates are compared with those presented in Javed et al.~\cite{Javed2017}.
This work presents a full palette twelve-component surrogate, as well as a reduced seven-component version.
It can be seen in Figure~\ref{F:FGIpionaa}, that the full surrogate has more \textit{n}-paraffins, aromatics, and naphthenes than FACE I, with less olefins and isoparaffins.
The reduced surrogate is a closer match for napthenes, \textit{n}-paraffins, and isoparaffins, but still a relatively high proportion of aromatics and no olefins.  
PIONA of the surrogate proposed by Javed et al.~\cite{Javed2017} match the PIONA of FACE I very well.
Figure~\ref{F:FGGcarbona} shows the \ce{C-C} groupings for the current surrogate being larger for carbon types 7 and 8 than FACE I, due to the high napthenes and aromatics.
Figure~\ref{F:FGIdista} shows our surrogates are generally out-performed by the Javed et al. surrogate across the majority of distillate temperatures.
We can see, in Figure~\ref{F:FGIfooa}, that our current surrogates have relatively high H/C and density errors, with less errors for RON and MON than the Javed et al. surrogate.  

Applying the FTIR-octane correlation to the Javed et al. FACE I surrogate, we predict a RON of 70.0 and MON of 69.8. These values are in better agreement to the target values than reported by Javed et al., who used the TRF linear-by-mol octane model.  When using our FTIR-octane model, the Javed et al. surrogate then outperforms our surrogates across all performance metrics. Indeed, the objective function is further minimized (0.59) over our proposed full (3.05) and reduced (1.8) surrogates. Clearly, our optimization routine did not find the global minimum of the objective function, suggesting our framework could benefit from additional parameterization of the species palette and initial mole fraction guesses---likely non-specific to FACE I.  Fortunately, this finding serves to validate the application of the FTIR-octane correlation, since the experimentally-validated surrogate of Javed et al. returned an objective function of nearly zero.

\subsubsection{FACE J}

\begin{table}[htbp]
\footnotesize
\centering
\begin{tabular}{@{}l c c c c@{}}
\toprule
Parameter 	& ~~FACE J\cite{Cannella:2014aa} & ~~Full~~ & ~Reduced~ & ~~Javed\cite{Javed2017} \\
\midrule
RON 	    & 73.8 & 72.7 &	73.2 & 70.6 \\
MON		    & 70.1 & 71.0 &	70.2 & 66.5 \\
Density [\si{\kilo\gram\per\m^3}] & 742  & 748 & 761 & 740  \\
H/C ratio   & 1.92 & 1.90 &	1.86 &	1.95  \\
\midrule
\% Volume distilled	& \multicolumn{4}{c}{Temperature [\si{\kelvin}]}  \\
\midrule
10			& 346& 352& 360&353 \\
20		    & 368& 360& 368&372 \\
30	        & 376& 368& 374&379 \\
40	        & 380& 375& 379&383 \\
50	        & 384& 381& 384&387 \\
60	        & 390& 388& 391&393 \\
70	        & 401& 400& 401&404 \\
80	        & 417& 422& 419&422 \\
\midrule
Species & \multicolumn{4}{c}{Molar \%} \\
\midrule
\textit{n}-butane   & & & & 10.5 \\
\textit{n}-pentane 	& & 3.0 &  & \\
\textit{n}-heptane  & & 21.8 & 23.7 & 24.5 \\
2-methylbutane	    & & 2.3 &  &  \\
2-methylpentane		& & 1.9 &  & \\
2-methylhexane	    & &  &  & 23.0\\
2,2,4-trimethylpentane	& & 18.1 & 16.4 & 12.0 \\
1-pentene	        & & 8.4 & 8.5 & \\
1-hexene	        & & 8.6 & 8.7 &  \\
cyclopentane	    & &  &  &  \\
cyclohexane         & & 8.5 & 12.3 & \\
toluene	            & & 3.4 & &  \\
\textit{o}-xylene   & & 5.0 & 15.0 & \\
1,2,4-trimethylbenzene	& & 15 & 15.4 & 30.0 \\
\bottomrule
\end{tabular}
\caption{
The full- and reduced FACE J surrogates compared with literature surrogates and the real FACE J properties. A blank entry indicates the species\slash parameter was not considered. A zero (0) indicates the species was in the palette, but not chosen by the optimizer.
}
\label{T:FACE_J}
\end{table}

\begin{figure}[htbp]
\centering
\begin{subfigure}[b]{.5\linewidth}
\includegraphics[width=85mm]{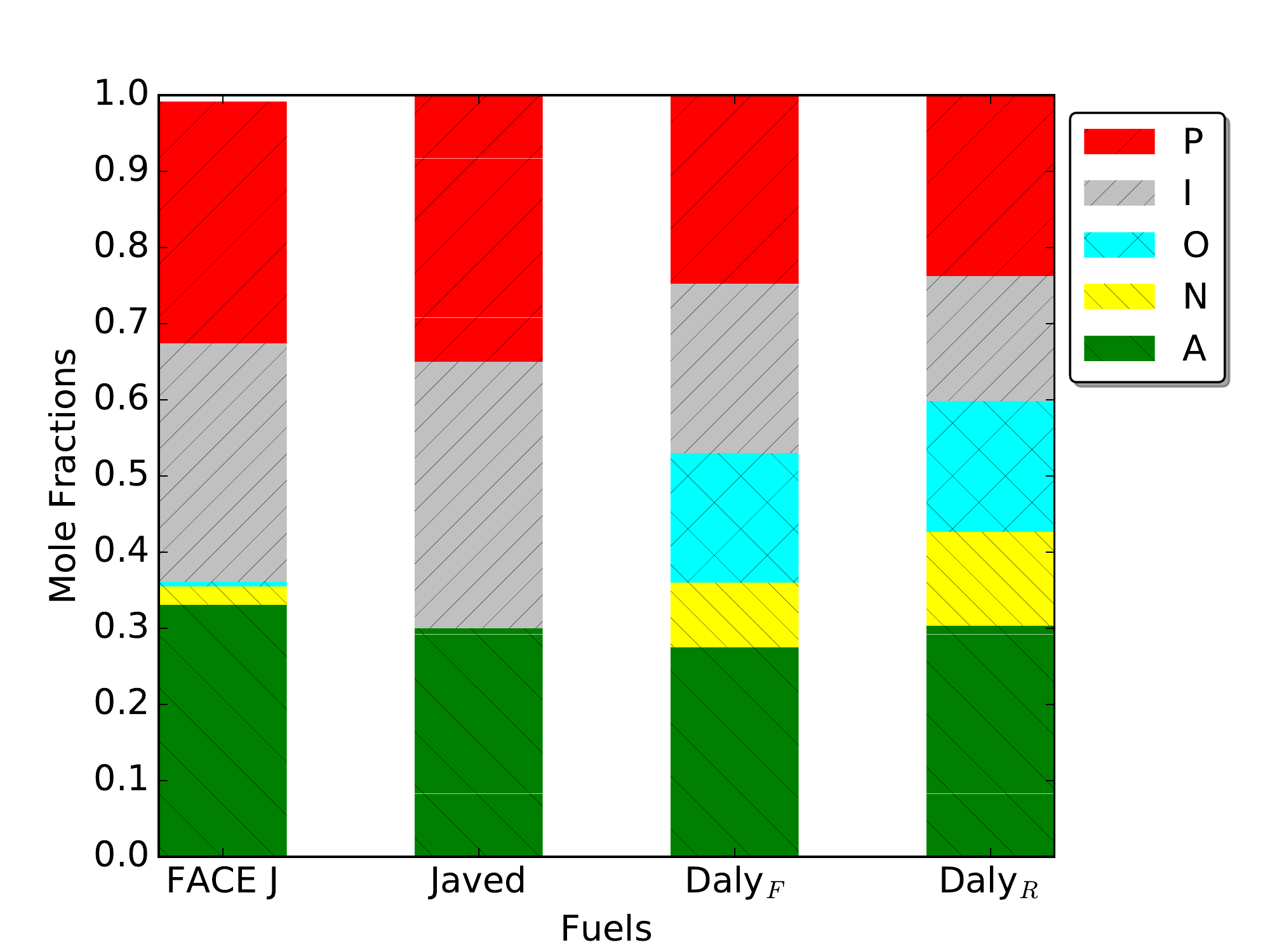}
\caption{Hydrocarbon class proportions}\label{F:FGJpionaa}
\end{subfigure}
~
\begin{subfigure}[b]{.45\linewidth}
\includegraphics[width=85mm]{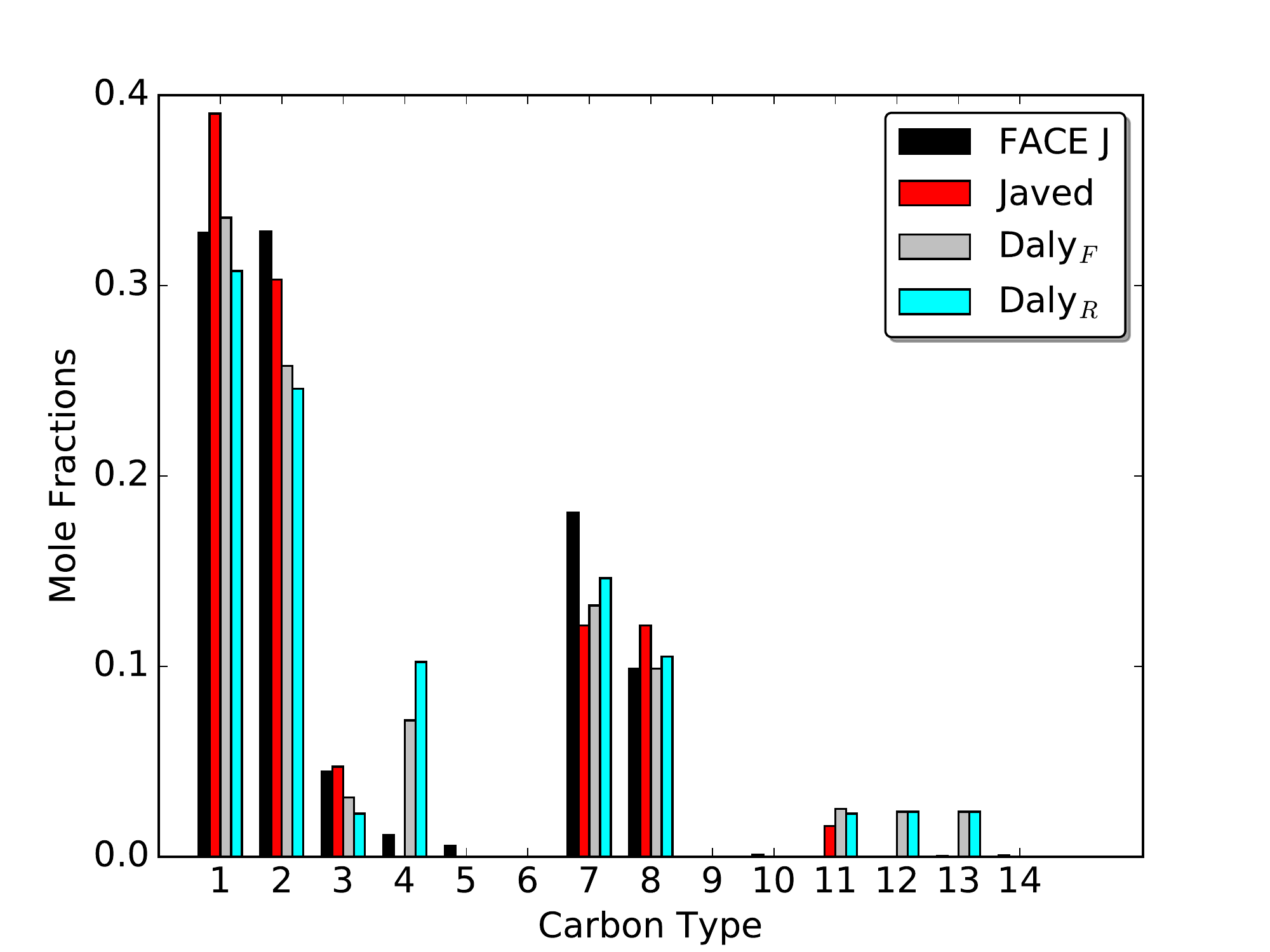}
\caption{C--C bond type proportions}\label{F:FGJcarbona}
\end{subfigure}
\\
\begin{subfigure}[b]{.5\linewidth}
\includegraphics[width=85mm]{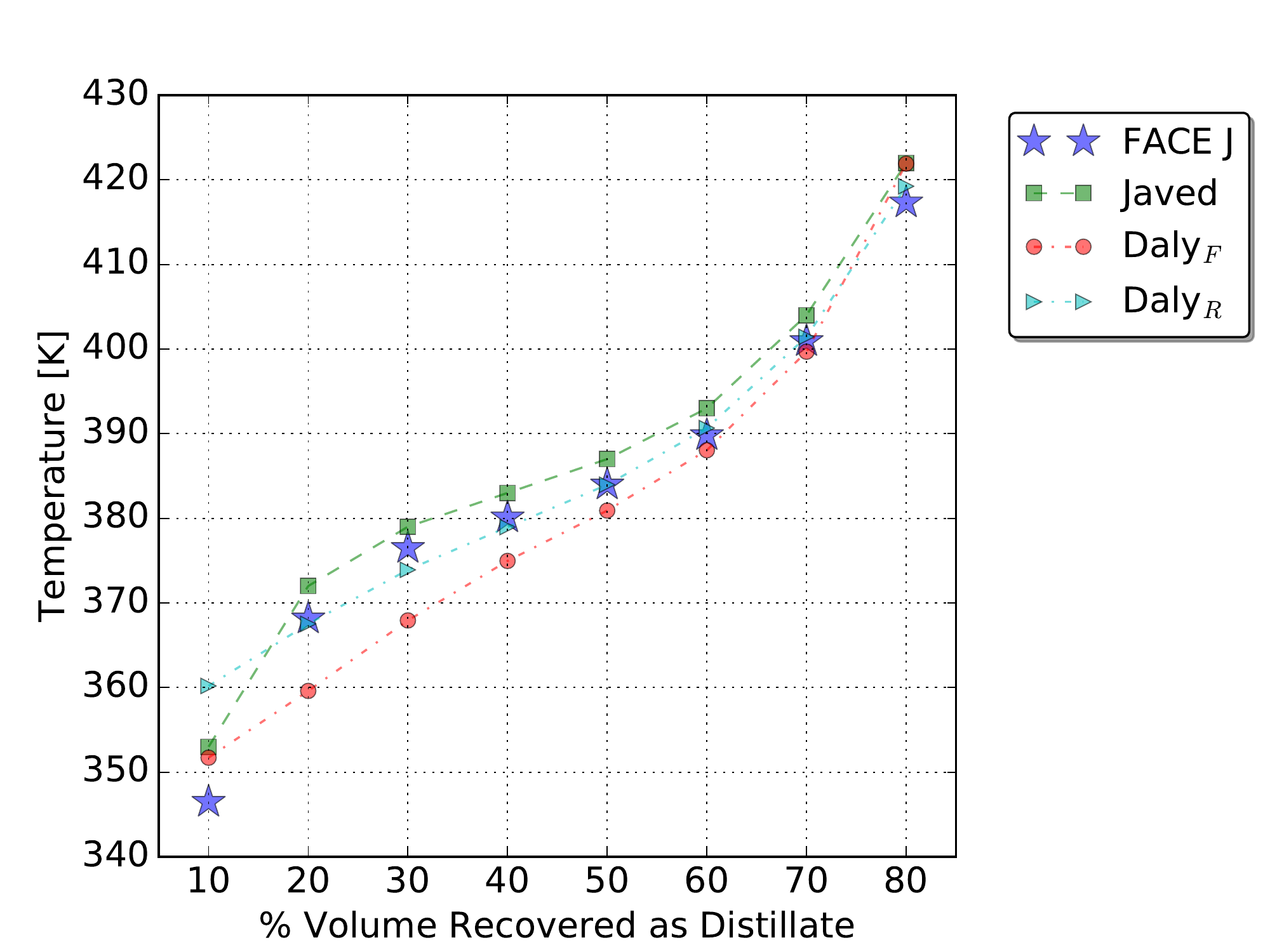}
\caption{Distillation Characteristics}\label{F:FGJdista}
\end{subfigure}
~
\begin{subfigure}[b]{.45\linewidth}
\includegraphics[width=85mm]{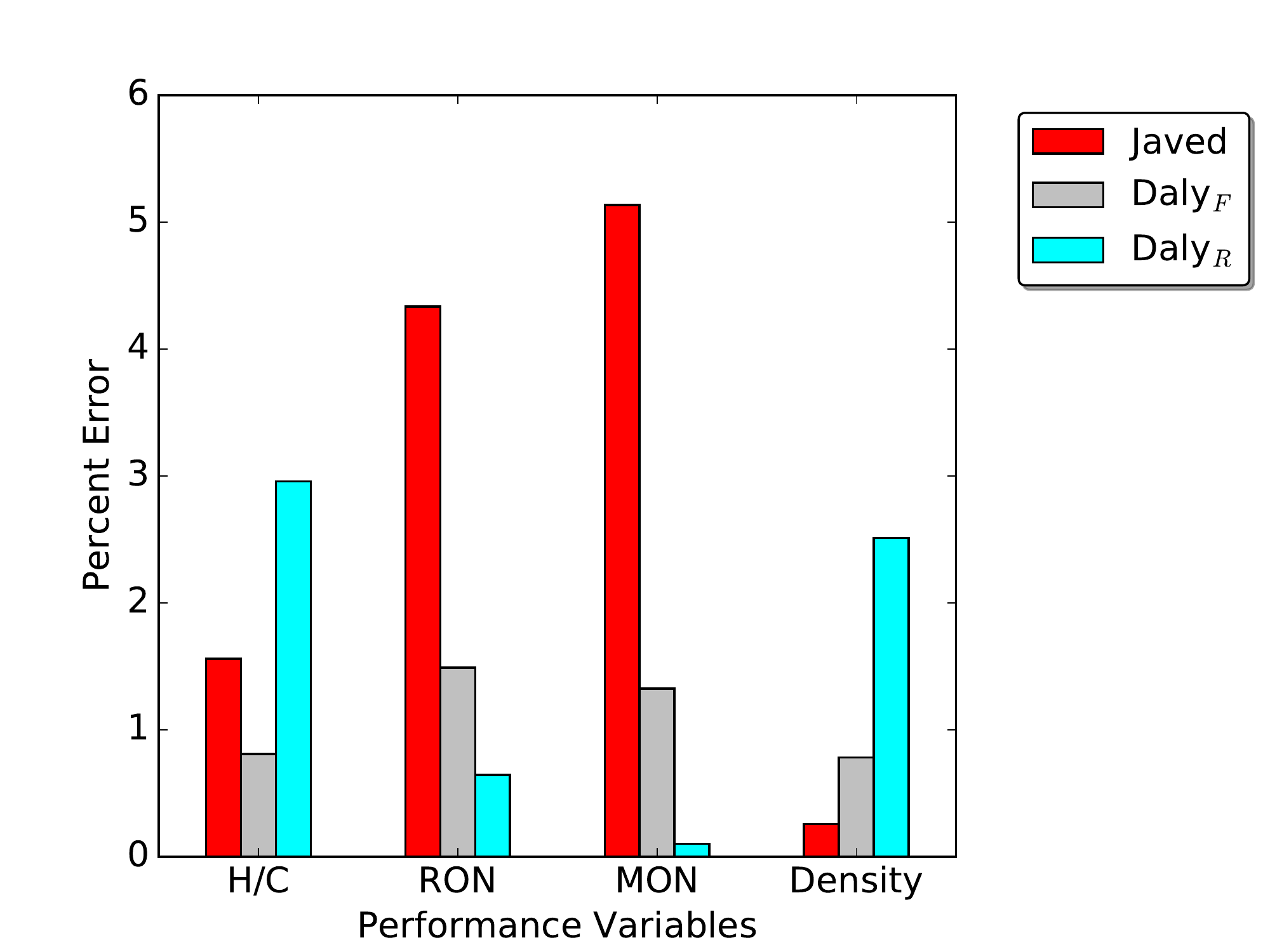}
\caption{Error in H/C, RON, MON, and density}\label{F:FGJfooa}
\end{subfigure}
\caption{Target property comparisons for FACE J and surrogates.  Shown are surrogates developed in this work and past literature efforts.}
\label{fig:face J properties}
\end{figure}

In Table~\ref{T:FACE_J}, FACE gasoline J surrogates are compared with the surrogate of Javed et al.~\cite{Javed2017}.
Our work presents a full palette eleven-component surrogate, as well as a reduced seven-component version.
It can be seen in Figure~\ref{F:FGJpionaa}, that the full surrogate has more olefins and napthenes than FACE J, with less isoparaffins and \textit{n}-paraffins; the reduced surrogate has similar PIONA to the full surrogate.
The Javed et al.~\cite{Javed2017} surrogate does not include the minor proportions of olefins and napthenes in FACE J, but matches the isoparaffins, \textit{n}-paraffins, and aromatics well.
Figure~\ref{F:FGJcarbona} shows the \ce{C-C} groupings for the current surrogates being relatively large for carbon types 4, 12, and 13 than FACE J due to the high olefins and napthenes, with a lower amount of carbon type 2, from lack of \textit{n}-paraffins and isoparaffins.
Figure~\ref{F:FGJdista} shows that our FACE J reduced surrogate T$_{b}$ values from 20 to \SI{80}{\percent} evaporated is similar to the Javed et al surrogate values, which match the target characteristics very well.  Our full surrogate tends to under-predict values across this distillate range. 
But we can see, in Figure~\ref{F:FGJfooa}, that our full component surrogate has lower H/C, RON, and MON errors, with higher density error.  On the other hand, our reduced surrogate has lower errors only for RON and MON.

%%%%%%%%%%%%%%%%%%%%%%%%%%%%%%%%%%%%%%%%%%%%%%%%%%%%%%%%%%%%%%%%%%%%%%%%%%%%%%%%%%%%%%%%%%%%%%%%%%%%%
\section{Conclusions}
\label{S:conclusions}
%%%%%%%%%%%%%%%%%%%%%%%%%%%%%%%%%%%%%%%%%%%%%%%%%%%%%%%%%%%%%%%%%%%%%%%%%%%%%%%%%%%%%%%%%%%%%%%%%%%%%
An existing gasoline surrogate formulation algorithm was further enhanced by incorporating novel chemometric models.
These models use attenuated total reflectance, Fourier transform infrared (ATR-FTIR) spectra of hydrocarbon fuels to predict research and motor octane numbers, alleviating the need for time-consuming auto-ignition simulations.
This work developed surrogates from a palette of 14 hydrocarbon species for the Fuels for Advanced Combustion Engine (FACE) gasolines. The palette includes candidate component species not previously considered in the literature: \textit{n}-pentane, 2-methylpentane, 1-pentene, cyclohexane, and \textit{o}-xylene.
Furthermore, reduction in the number of components in surrogates was automated based on a mole fraction threshold.  As such, the ability to match the properties of the 10 FACE gasoline was evaluated.

This technique yields surrogates for gasoline fuels that accurately match target properties.
On average, our ``full'' (7--12 species) and ``reduced'' palette (4--7 species) surrogates match all the target properties of the FACE gasolines within \SI{5}{\percent}. 
RON, MON, and distillation curves were matched within \SI{1}{\percent}, with H/C and density within \SI{2.5}{\percent} and \SI{4.8}{\percent}, respectively. FACE G presented the only challenge to create a surrogate for. This seems to be due to the high octane sensitivity (S=11) coupled with low H/C (1.83) and high density (\SI{760}{\kilo\gram\per\meter^3}), attributable to low olefins with high aromatics and paraffins present in this fuel.  This was a challenge due to three contributing factors: 1) the optimization routine not being strictly constrained to follow hydrocarbon class proportions to match the target fuel, 2) the IR-octane models converging to high olefinic contents in order to formulate a high octane sensitivity fuel, and 3) the objective function highly weighting octane more-so than H/C or density.
In short, the optimization routine blended a high sensitivity fuel (by way of high olefinic content), at the expense of properly matching H/C, density, and hydrocarbon class proportions.  We conclude that the formulation framework utilizing the IR-octane models, in their current state, are not completely adequate to match fuels having a combination of sensitivity greater than 10, H/C ratio greater than 1.8, and density less than \SI{760}{\kilo\gram\per\meter^3}---in short, low olefinic-content fuels with sensitivity over 10.
With that said, the other nine FACE gasoline surrogates adequately matched all target properties.  We suggest using the ``full'' palette surrogates since PIONA proportions are better met with these surrogates, in addition to the other target properties. However, if a study requiring minimal computational expense is desired, the formulated ``reduced'' palette surrogates can be used (at the expense of matching PIONA).  

In some cases, the ``reduced'' surrogates have a lower objective function value, indicating a more optimal fuel mixture over the ``full'' surrogate.  This result is counter intuitive. We expect the objective function to be zero as the components in the surrogate palette approach those contained in target fuel, and also blended in the correct proportions by the optimization routine.
We suggest that modeling artifacts cause the ``reduced'' surrogates to outperform the ``full''. The overall accuracy of predicted fuel properties could be reducing as the species palettes grows.  Larger species palettes also bring the increased possibility that the optimization routine is not guaranteed to return a global minimum.  We found this to be the case with FACE I, where the  established surrogate from Javed et al.~\cite{Javed2017} provided a lower objective function value over both our formulated surrogates. Because of this, the fuel palette selection logic should be revisited.  Additionally, weighting factors for the objective function may also need to be determined on a per-surrogate basis, due to the large variability in the species palette that could influence parameter sensitivity.  These intricacies and their impact on modeling results were not investigated in this work, and should be considered in future efforts.       

The surrogates created in this work were compared to literature~\cite{Sarathy2014,Ahmed2015,Sarathy2016,Javed2017}.
Our surrogates, on average, better-match RON, MON, and distillation characteristics at 0.46, 0.65, and \SI{0.93}{\percent} error, respectively, with literature surrogates at 1.2, 1.1, and \SI{1.8}{\percent} error. Although, we worse-match density and hydrogen-to-carbon ratio at 3.31 and \SI{6.81}{\percent} error with literature surrogates at 1.3 and \SI{2.3}{\percent}. We also find that our molar quantities of carbon--carbon bond types deviate at 2.66 molar\% with literature at 1.9 molar\%. 

Surrogates from the literature used vastly different approaches to predict RON, MON, or S.
Those approaches were either computationally expensive, not valid for the hydrocarbons considered in this work, or not designed with the intent to predict fuels where molecule-molecule interactions are more prevalent---the methodology in this work simultaneously minimized computational effort and is applicable over the wide range of fuels considered.
We predicted the RON and MON values with our FTIR-octane models for three proposed surrogates from literature including FACE F, G and I; these do not include n-butane in the species palette, a restriction our octane model requires. 
For the LLNL FACE F and G surrogates~\cite{Sarathy2016}, RON and MON were found to have large discrepancies between our calculated values to those provided by the computationally-heavy, ignition delay to octane correlation.
The FACE I surrogate of Javed et al., with RON and MON based on the TRF linear-by-mol blending formula, agreed with our FTIR-octane predictions when evaluating this surrogate. Further investigations are warranted to verify the foresight of our FTIR-octane model more specific to low-component surrogate fuels seen during optimization.  In this manner, a definitive conclusion could be made as to which octane model performs best in predicting surrogate mixtures.      

The surrogates proposed in this work ultimately need to be validated by experimental efforts, such as those in literature~\cite{Sarathy2014,Ahmed2015,Sarathy2016,Javed2017}.  Based on the attractive results for many of the generated surrogates in this work, we aim to follow up with experimental validation efforts.      
The altered gasoline surrogate formulation framework generates surrogates in an expedited, and possibly a more accurate manner; it should be considered for further refinement and adoption.
Alternately, this methodology could be extended to formulating diesel and jet-fuel surrogates.
It should be possible to extend the methodology of Daly et al.~\cite{Daly:2016} to create an IR-cetane number model to use in conjunction with the formulation framework.  

%%%%%%%%%%%%%%%%%%%%%%%%%%%%%%%%%%%%%%%%%%%%%%%%%%%%%%%%%%%%%%%%%%%%%%%%%%%%%%%%%%%%%%%%%%%%%%%%%%%
\section*{Acknowledgments}
%%%%%%%%%%%%%%%%%%%%%%%%%%%%%%%%%%%%%%%%%%%%%%%%%%%%%%%%%%%%%%%%%%%%%%%%%%%%%%%%%%%%%%%%%%%%%%%%%%%

The authors gratefully acknowledge the Chevron Energy Technology Company for supporting this research, as well as Dr.~Marcia Huber for her expert guidance in distillation calculations.

\section{Supporting Information}

\begin{description}[labelsep=1em, align=left, labelwidth=1em,labelindent=0em]
{\setstretch{0.25}
\item[ FACE Surrogates.pdf] Plots illustrating properties of FACE gasolines A--J for ``full'' and ``reduced'' surrogates vs. the measured properties of the fuels  
\item[RON model.xls] Spectral weightings ($W_{\nu}$) and offset ($b$) for IR-RON model (Eq.~\ref{eq: octane relation}) 
\item[MON model.xls] Spectral weightings ($W_{\nu}$) and offset ($b$) for IR-MON model (Eq.~\ref{eq: octane relation})
}
\end{description}

\bibliography{refs}

\end{document}